\documentclass[10pt,twocolumn]{IEEEtran}

\usepackage{cite}
\usepackage{graphicx}
\usepackage{graphics}
\usepackage{amsmath}
\usepackage{amssymb}
\usepackage{algorithm}
\usepackage{algorithmic}
\usepackage{array}
\usepackage{epsfig}
\usepackage{color}
\usepackage{multirow}
\usepackage{booktabs}
\usepackage{float}

\newtheorem{lemma}{Lemma}
\newtheorem{proposition}{Proposition}
\newtheorem{definition}{Definition}
\newtheorem{theorem}{Theorem}

\usepackage{subcaption}
\usepackage{xcolor}
\usepackage{hyperref}
\usepackage{mathtools}

\newcommand{\figref}[1]{Fig.~\ref{#1}}

\begin{document}

\title{Unraveling the Viral Spread of Misinformation: Maximum-Likelihood Estimation and Starlike Tree Approximation in Markovian Spreading Models}

\author{
 \IEEEauthorblockN{Pei-Duo YU and Chee~Wei~Tan$^*$\\}
  \IEEEauthorblockA{Chung Yuan Christian University, Nanyang Technological University$^*$\\
 peiduoyu@cycu.edu.tw, cheewei.tan@ntu.edu.sg$^*$}
}

\maketitle

\begin{abstract}
Identifying the source of epidemic-like spread in networks is crucial for removing internet viruses or finding the source of rumors in online social networks. The challenge lies in tracing the source from a snapshot observation of infected nodes. How do we accurately pinpoint the source?  Utilizing snapshot data, we apply a probabilistic approach, focusing on the graph boundary and the observed time, to detect sources via an effective maximum likelihood algorithm. A novel starlike tree approximation extends applicability to general graphs, demonstrating versatility. Unlike previous works that rely heavily on structural properties alone, our method also incorporates temporal data for more precise source detection. We highlight the utility of the Gamma function for analyzing the ratio of the likelihood being the source between nodes asymptotically. Comprehensive evaluations confirm algorithmic effectiveness in diverse network scenarios, advancing source detection in large-scale network analysis and information dissemination strategies.
\end{abstract}

\section{Introduction}

Epidemic-like spreading represents a crucial topic in network science, extensively explored in the existing literature \cite{lf,science,pnas}. Specifically, the proliferation of malicious information in networks has emerged as a significant cybersecurity challenge \cite{Wasserman94}. Detecting the sources of malicious information has numerous applications, such as eradicating computer viruses or identifying the origins of rumors spreading on the Internet or online social networks. The COVID-19 pandemic marked a unique global crisis, intertwining the epidemics of the virus with an overwhelming surge of misinformation. It became the first pandemic in history that triggered an {\it epidemic of online misinformation}, which significantly undermined the efficacy of online social networks and disrupted public health risk communications \cite{naturemisinformation,infodemic,tanyusurvey}. The World Health Organization (WHO) swiftly declared war against the COVID-19 Infodemic, referring to the viral spreading of pandemic-related misinformation or disinformation on social media \cite{infodemic}. 

\subsection{Source Detection Problem}
The problem of identifying the origin of contagion was initially explored in the seminal work \cite{Shah2011,Shah2012}. The task at hand focuses on identifying the origin of a spreading event, given a single snapshot depicting the network of connections among individuals labeled as ``infected," based on the Susceptible-Infectious (SI) model described in \cite{Bailey1975} for contagion propagation. For simplicity, we will refer to nodes that have acquired the malicious information as {\it infected nodes} and those that have not as {\it susceptible nodes} or as {\it uninfected nodes}. This problem is framed as a graph-constrained maximum likelihood estimation, intensified by the interaction between spreading dynamics and network topology, crucial for developing scalable, efficient algorithms, as highlighted in \cite{tanyusurvey}. The concept of rumor centrality was first introduced in \cite{Shah2011,Shah2012} as a novel method for addressing the challenge of maximum likelihood estimation in the context of rumor spread analysis. This approach involves assigning a numerical value to each node within the network exposed to the rumor to identify the node with the highest value, designated as the {\it rumor center}. Specifically, the rumor center is the optimal solution for the maximum likelihood estimation problem for graphs embedded in infinite regular trees. The rumor center can be efficiently computed using message-passing algorithms, as described in\cite{mackay}. Moreover, the effectiveness of this strategy can be evaluated asymptotically, particularly as the number of nodes affected by the rumor escalates significantly \cite{Shah2011,fuch}.

The maximum likelihood estimation for general network topology remains unsolved, with only a few special cases having optimal solutions. These cases involve infinite underlying graphs with specific structures, such as degree-regular tree graphs \cite{Shah2011} or starlike graphs \cite{Srikant_ICASSP2016}, or special scenarios with finite underlying graphs \cite{pd_ASONAM, tanyusurvey}, or graphs containing cycles \cite{pd_ITW2017,tanyusurvey}. However, some heuristics based on network centrality, such as those in \cite{Shah2011,Shah2012, pd_jstsp2022}, have demonstrated good performance. Since the initial work by \cite{Shah2011,Shah2012}, the problem formulations have undergone several modifications. For instance, generalizations to random trees were explored in \cite{fuch}, extensions with suspect sets were introduced in \cite{tan1}, and the problem was extended to multiple source detection in \cite{tay}. Furthermore, an extension for detection with multiple snapshot observations was studied in \cite{tan2}. A comprehensive survey on this topic can be found in \cite{tanyusurvey}.

\subsection{Network Immunization Problem}
In contrast to the source detection problem, the network immunization problem has attracted significant attention in the current literature \cite{DAVA, MCWDST, pd_JSTSP2018, CONTAIN, netshield2016, sparseshield}. Network immunization focuses on strategies to protect a network from arbitrary diffusion, such as misinformation, disinformation, and harmful content, by detecting and immunizing key nodes within the network. Several methods have been developed to tackle this problem effectively. For instance, DAVA \cite{DAVA} proposes a strategy that first merges all infected nodes as a super node and inputs the resultant graph to construct a dominator tree, identifying optimal nodes for immunization. This tree-based approach is particularly effective in large-scale networks, as it targets the most influential nodes while minimizing the number of interventions. Similarly, the MCWDST paper \cite{MCWDST} introduces a minimum-cost weighted directed spanning tree (MCWDST) algorithm for real-time fake news mitigation, balancing the cost and effectiveness of interventions. It leverages deep learning architectures to detect fake news and strategically selects nodes for immunization based on their harmfulness ranking within the network.

Other notable proactive immunization strategies include NetShield \cite{netshield2016} and SparseShield\cite{sparseshield}. These approaches are grounded in the spectral properties of the adjacency matrix $A$ of the graph. NetShield computes the vulnerability of a network based on the dominant eigenvalue of $A$, constructing a priority queue of nodes for immunization within a given budget. SparseShield extends this by incorporating a priority multiplier and creating an ordered ranking list for effective resource allocation. Community-based methods such as CONTAIN \cite{CONTAIN} and ContCommRTD\cite{CCRTD} take a different approach by focusing on the community structure of networks. CONTAIN leverages community detection algorithms to identify and prioritize node sets for immunization, often outperforming spectral-based methods. ContCommRTD further utilizes community properties in social networks to design a real-time reporting system for hazard-related events.

Building upon the insights from existing immunization strategies that focus on intervention after the detection of harmful content, our work offers a complementary perspective by aiming to identify the source of the spread. By utilizing the ML estimation, we provide a method to enhance the effectiveness of strategies like NetShield, SparseShield, CONTAIN, and ContCommRTD. Specifically, accurate source detection can inform these strategies where to allocate resources effectively, thereby improving overall network immunization. For example, in a social network where fake news is spreading, identifying the initial source of the fake news helps platforms or organizations target users (nodes) who are most likely to further disseminate the misinformation. These users can then be prioritized for fact-checking or information blocking, effectively stopping the spread at its origin.

\subsection{Our Approach and Main Contributions}
We consider a probabilistic approach to rumor source detection using a continuous-time SI model that shares similarities with the discrete-time SI model in \cite{Shah2011,Shah2012}. Both models share the feature that the spread time between nodes is exponentially distributed with a constant mean. Other work employing similar models include \cite{pd_2024_CISS,icassp2018, kdd2019}. The fundamental concept of this paper is to define the {\it rumor boundary} of the observed rumor graph, facilitating a probabilistic method for estimating the source. Specifically, source estimation can be effectively determined by a message-passing algorithm on tree networks. The use of rumor boundary introduces a unique characterization method distinct from the centrality-based approaches introduced in prior works \cite{Shah2011, Shah2012, pd_JSTSP2018,pd_jstsp2022}. The initial research presented in this paper was previously published in \cite{zheng1}. We summarized the main contributions of this paper as follows:

\begin{itemize}
\item We present a detailed probabilistic examination focusing on the boundary of the rumor graph. By analyzing the properties of graph connectivity and the effects of observation timing, we provide novel insights into the dynamics of rumor propagation in networked structures.

\item 
We formulate the source detection problem as a maximum likelihood estimation by leveraging observed graph data and precise observation timings. We show that when the underlying network is a degree regular tree, the ratio of the likelihood of being the source between two nodes is independent of the observed time $T$. We propose an effective message-passing algorithm tailored for tree networks, establishing its optimality for starlike trees.

\item  To address the challenge of maximum likelihood estimation for source detection in general graphs, we propose a novel starlike tree approximation for general graphs and then demonstrate the algorithm's performance in graphs with cycles. This extends its utility beyond tree networks, showcasing its versatility and robustness.

\item Our formulation provides valuable insights into network resilience under complex systems at vast scales. In particular, we show that the asymptotics of gamma functions offer a potent tool for analyzing graph-theoretic features in large networks. Leveraging their properties enables the approximation of essential combinatorial quantities, facilitating the characterization of source estimation.

\item Our comprehensive performance assessment illustrates that the proposed algorithm achieves robust numerical results across different random graphs, including those with cycles and intricate boundaries.
\end{itemize}

In our study, we consider the same spreading model as in \cite{Shah2011}. The analysis in \cite{Shah2011} is simplified by leveraging the memoryless property of the exponential distribution, which allowed the analysis to focus primarily on the structural properties of the network without considering the specific observed time $T$. In contrast, our approach deliberately incorporates the observed time into the analysis. By taking the observed time into account, we gain deeper insights into the dynamics of rumor spreading and improve the accuracy of source detection, particularly in non-regular networks. The work in \cite{Mahmoud_2020} also emphasizes the importance of time by investigating how long it takes for a certain proportion of the community to become spreaders. By accounting for the time taken for the spread, our model can better capture the progression of rumors spread across networks, making the estimation of the source more precise. This temporal dimension adds depth to our analysis, distinguishing our approach from models that focus solely on structural properties, thereby enhancing the relevance and applicability of our method in real-world scenarios. Notably, when the underlying network is a regular tree, our method yields the same results as those obtained in \cite{Shah2011}, reflecting the inherent time independence of likelihood ratios in such symmetric structures. This demonstrates that our approach encompasses results in the literature and provides additional value by incorporating the temporal dimension, making it applicable to a wider network topologies.

Collectively, these contributions mark a significant advancement in the field of rumor source detection, offering valuable insights and methodologies that transcend traditional boundaries. This research paves the way for enhanced network analysis and information dissemination strategies.

The organization of this paper is as follows: Section \ref{sec:model} details the SI spreading model utilized in this study, incorporating an observed time variable $T$, and introduces the ML estimator for identifying the source. In Section \ref{sec:tree}, we show that the likelihood ratio between two nodes in this temporal SI spreading model is independent of the observed time when the underlying network is a regular tree. Next, we analyze the likelihood of a specific node within the context of the rumor boundary and proceed to derive the ML estimator for tree graphs. In Section \ref{sec:starlike}, we propose a starlike tree approximation for the source estimator in general graphs and analyze the likelihood ratio asymptotically. Section \ref{sec:sim} assesses the effectiveness of our probabilistic method for source estimation across various network structures. We further discuss other possible spreading distributions in Section \ref{sec:discussions}. The paper is summarized in Section \ref{sec:conclusion}.

\section{System Model}\label{sec:model}
In this section, we outline the rumor-spreading model and introduce a maximum likelihood (ML) estimator for identifying the source, focusing specifically on the impact of time on infected nodes, diverging from the approach used in \cite{Shah2011,Shah2012}.

\subsection{Rumor Spreading Model}
\label{sec:spreading}
In our study, the network is represented by an undirected graph $G=(V(G), E(G))$, with $V(G)$ being the node set of $G$ and $E(G)$ comprising edges of the form $(v_i, v_j)$, for $v_i, v_j \in V(G)$. The degree $d_i$ of node $v_i$ refers to the count of adjacent neighbors of $v_i$. We use $d(v_i,v_j)$ to denote the distance (number of hops) between $v_i$ and $v_j$ in a graph. If $v_i$ and $v_j$ are defined in multiple graphs, we use $d_G(v_i,v_j)$ to represent the distance in a specific graph $G$ \cite{graph_bolo,graph_west}. The dynamics of rumor dissemination are represented through the susceptible-infected (SI) model, a simplification of the SIR model for disease transmission \cite{Bailey1975}. In the SI model, each node in the network can be in one of two states:
\begin{itemize}
    \item Susceptible (S): A node is susceptible if it has not yet been infected (or received the rumor).
    \item Infected (I): A node is infected if it has received the rumor and can spread it to its neighbors.
\end{itemize}
Once a susceptible node comes into contact with an infected node, it has a chance to become infected. Once a node becomes infected, it remains infected for the rest of the process. There is no recovery state, as the goal of the model is to track the spread of information over time without recovery. We use the notation $\textbf{v}_i$ to denote the event that $v_i$ is infected; conversely, ${\bar{\textbf{v}}_i}$ denotes the event that $v_i$ is not infected. In practical situations, constructing the rumor subgraph $G_N$ requires specific methods. For example, if we are modeling the spread of a virus, $G_N$ can be obtained through contact tracing conducted by public health authorities. Contact tracing identifies individuals who have been in close contact with an infected person, forming the subgraph of potential transmissions. If the focus is on the dissemination of fake news in social networks, the detection of $G_N$ involves more complex approaches based on deep learning architecture. This task can be tackled using methods discussed in recent research papers. For instance, using transformer-based models for misinformation detection \cite{Elena2022_math, trumorgpt}, combining BART and RoBERTa to classify news articles. Similarly, the work \cite{Elena2023_math} employs document embeddings and machine learning models, such as Na\"ive Bayes, perception or LSTM,  to detect fake news articles. Another approach \cite{Elena2024_KBS} proposes a deep neural network ensemble method that integrates both textual and social contexts for fake news detection. Lastly, the work \cite{elena2021_ACC} utilizes various deep learning architectures, including RNNs, LSTMs, and CNNs, to classify news into multiple categories using word embeddings. These methods collectively highlight how the construction of $G_N$ can be approached differently depending on the type of spread being studied, whether it involves a virus or misinformation on social networks.

Next, we describe the rumor-spreading process on a given networked structure. At time $t=0$, the spreading process starts with a single infected node (the rumor source) denoted as $v^{\star}$. As time progresses, the infected node spreads the infection to its neighboring susceptible nodes, and these nodes then spread the infection further to their neighbors. The time $\tau_{ij}$ takes for a node $v_i$ to infect node $v_j$ is modeled as a random variable following an exponential distribution with a rate parameter $\lambda$. The infection continues to spread through the network until all nodes in the connected component have been infected or the observed time $T$ is reached. This implies that the infection process is stochastic, and the time $\tau_{ij}$ between the moment node $v_i$ becomes infectious and when it infects node $ v_j$ is not deterministic but varies according to the exponential distribution. 
The probability density function of $ \tau_{ij}$ is given by:

\[
f(\tau_{ij}) = \lambda e^{-\lambda \tau_{ij}}, \quad \text{for } \tau_{ij} \geq 0.
\]
Specifically, $\lambda$ represents the expected number of infections per unit time. For example, if time is measured in hours, then with $\lambda =2$, on average, there is one infection every $30$ minutes. For simplicity, we assume that each infection event is independent, with the same rate $\lambda=1$ for all pairs of nodes.


\subsection{Rumor Source Estimator}
Assuming a rumor starts spreading from a node $v^\star$ in a given network $G$ at time $t=0$. By the time $t=T$, we observe $G$ and identify $N$ nodes as infected, forming a connected subgraph of $G$. We call $G$ the underlying graph and $G_N$ the rumor graph. Our objective is to propose a source estimator $\hat{v}$ for deducing the origin of the rumor by considering the network structure and the observed time $T$. Utilizing Bayes' theorem, the ML estimator for $v^\star$ identifies the node that maximizes the probability of correct detection, as determined by
\begin{equation}\label{eq:ml}
\hat{v}\in\arg\max_{v\in V(G_N)} P(G_N\text{ is observed}\,|\,v=\text{source}, T),
\end{equation}
where $P(G_N\text{ is observed}\,|\,v=\text{source}, T)$ represents the likelihood of observing the rumor graph $G_N$ at time $T$, given $v$ as the presumed source of the rumor. For simplicity, we denote the likelihood $P(G_N\text{ is observed}\,|\,v=\text{source}, T)$ as $P(G_N| v, T)$.
It should be noted that in the event of ties within the solution of the maximum likelihood estimation, they are resolved uniformly at random, indicating that the solution might not be singular.


\section{Rumor Source Detector for Tree Graphs}\label{sec:tree}

\begin{figure}
    \begin{minipage}{.5\linewidth}
        \centering
\includegraphics[scale=0.3]{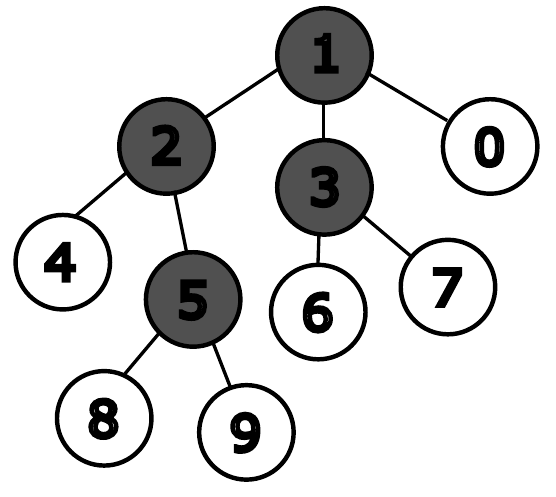}
    \end{minipage}%
    \hfill
    \begin{minipage}{.5\linewidth}
          \begin{tabular}{|c|c|}
        \hline
        $v_i$ & $P(G_4|v_t,T)$  \\
        \hline
        $v_1$ &  $\frac{1}{2}\cdot e^{-6T}\cdot (e^T-1)^3 $ \\ \hline
        $v_3$ &  $\frac{1}{6}\cdot e^{-6T}\cdot (e^T-1)^3$ \\ \hline
        $v_2$ & $\frac{1}{2}\cdot e^{-6T}\cdot (e^T-1)^3$  \\ \hline
        $v_5$ & $\frac{1}{6}\cdot e^{-6T}\cdot (e^T-1)^3$ \\\hline
          \end{tabular}
    \end{minipage}
\caption{An example rumor graph where the underlying graph is a $3$-regular tree. The figure illustrates infected nodes as grey and susceptible nodes as white circles. It is important to note that the underlying graph could potentially extend infinitely, but this detail is omitted due to space constraints. The likelihood of each vertex being the source that leads to the observed rumor graph $G_4$ is listed in the table on the right.}
\label{fig:treeEg}
\end{figure}

This section evaluates $P(G_N|v,T)$ in tree graphs. We need to compute a multiple integral with more than $N-1$ variables when computing $P(G_N|v,T)$ for an $N$-nodes tree. We take Fig. \ref{fig:treeEg} for example; the value of $P(G_4|v_1,T)$ can be obtained by calculating a multiple integral with nine variables since the graph has nine edges. To present the forthcoming theorem, we define the concept of a $d$-regular tree. A $d$-regular tree is defined to be a tree graph where each node has $d$ neighboring nodes, hence a $d$-regular tree has infinite number of nodes. Here, we consider finite-size regular trees, where only leaf nodes may not have $d$ neighbors. When the underlying network $G$ is a regular tree and $G_N$ does not contain any leaf node in $G$, the ratio of $P(G_N|v,T)$ between any two nodes is independent of the time $T$. In the following, we consider an example of $G$ as a $3$-regular tree and $G_4\subset G$ is the observed rumor graph shown in Fig. \ref{fig:treeEg}.  

We can observe that the first constant is the only difference between $P(G_4|v,T)$ of any two nodes in $G_4$. The constant comes from the result of multiple integrations. To further clarify, we suppose at time $t = 0$, a node in graph $G$ starts spreading a rumor. By the time $t = T$, we observe that three nodes, $v_1, v_2, v_3, v_5$, have learned the rumor. The induced subgraph formed by $v_1, v_2, v_3, v_5$ is the rumor graph $G_N$. Based on the assumption that the time $\tau_{ij}$ takes for a node $v_i$ to infect node $v_j$ is modeled as a random variable following an exponential distribution with a rate parameter $\lambda=1$. We can compute the probability $P(G_N|v_1,T)$ by independently considering three branches of the root note $v_1$. That is, $P(G_N|v_1,T)=P(\textbf{v}_2\cap\textbf{v}_5\cap{\bar{\textbf{v}}_4}\cap{\bar{\textbf{v}}_8}\cap{\bar{\textbf{v}}_9}|v_1,T)\cdot P(\textbf{v}_3\cap{\bar{\textbf{v}}_6}\cap{\bar{\textbf{v}}_7}|v_1,T)\cdot P({\bar{\textbf{v}}_0}|v_1,T)$. For example, the likelihood $P(\textbf{v}_3\cap{\bar{\textbf{v}}_6}\cap{\bar{\textbf{v}}_7}|v_1,T)$ can be computed as follows:

\begin{align*}
&P(\textbf{v}_3\cap{\bar{\textbf{v}}_6}\cap{\bar{\textbf{v}}_7}|v_1,T)\\
&=\int_{0}^{T} \int_{T-\tau_{1,3}}^{\infty}  \int_{T-\tau_{1,3}}^{\infty} e^{-\tau_{1,3}-\tau_{3,6}-\tau_{3,7}} \,d\tau_{3,6}\,d\tau_{3,7}\,d\tau_{1,3}.  
\end{align*}
We can compute $P(\textbf{v}_2\cap\textbf{v}_5\cap{\bar{\textbf{v}}_4|v_1,T}$ and $P({\bar{\textbf{v}}_0}|v_1,T)$ in a similar way. Thus, by multiplying three likelihood together, we have $P(G_N|v_1,T)=\frac{1}{2}\cdot e^{-6T}\cdot (e^T-1)^3$. Based on the observation, we present some theoretical results in a $d$-regular tree in the following, which coincides with the results in \cite{Shah2011}.

\renewcommand{\arraystretch}{1.5}

\begin{lemma}
\label{lem:T_inde}
  Let $G$ be a $d$-regular tree, and let $G_N\subset G$ be the observed rumor graph of $G$ with no leaf node of $G$, under the assumption that the infection spreads according to the SI model as described in Section \ref{sec:spreading}. For each $v\in V(G_N)$, the likelihood $P(G_N|v,T)$ must have the following form 
  \begin{equation}
  \label{eq:d_reg_close_form}
      P(G_N|v,T)=k\cdot e^{-(N(d-2)+2)T}\cdot (e^{(d-2)T}-1)^{N-1},
  \end{equation}
  for $N\geq 2$, and $k$ is a number independent from $T$.
\end{lemma}
In Lemma \ref{lem:T_inde}, the term $(N(d-2)+2)$ in the power of the exponential means there are $(N(d-2)+2)$ uninfected neighbors of $G_N$ in $G$ and the term $N-1$ means there are $N-1$ infected edges, i.e., the edge connecting two infected nodes. For example, if $G_N$ is on a $3$-regular tree with $N=4$, then for all $v\in V(G_N)$, we have $P(G_N|v_i,T)=k_i\cdot e^{-6T}(e^{T}-1)^{3}$, where $k$ is a scalar as shown in Fig. \ref{fig:treeEg}. To present the next result, we define $g^v_u\subset G_N$ to be the subtree rooted at $u$ when we treat $G_N$ as a rooted tree with root $v$, and we denote $|g^v_u|$ the number of nodes in the subtree $g^v_u$.

\begin{theorem}
\label{thm:like_ratio_reg}
    Let $G$ and $G_N$ be defined as in Lemma \ref{lem:T_inde}. Then for any two adjacent nodes $u$ and $v\in V(G_N)$, we have 
    \begin{equation*}
        \frac{P(G_N|v,T)}{P(G_N|u,T)}=\frac{|g^u_v|}{N-|g^u_v|}. 
    \end{equation*}
\end{theorem}

The above Theorem \ref{thm:like_ratio_reg} shows that when $G_N$ is a subgraph of a regular tree, the ratio of the likelihood being the source between two adjacent nodes in $G_N$ is independent of the time $T$ and only related to the graph topology of $G_N$. Hence, we can find the maximum likelihood estimator for the source on $G_N$ by only considering the structure of $G_N$, which leads to the same result in \cite{Shah2011}. That is, the ratio of the likelihood of being the source for two neighboring nodes can be derived from the structure of $G_N$ which matches Equation (12) in \cite{Shah2011}.

Next, we consider the case of general trees. Our goal is to approximate $P(G_N|v,T)$ by considering the nodes on the rumor graph boundary, and this also enables a message-passing algorithm. To analyze the {\it boundary} of the rumor graph, we first define the $d(v,G_N)=\min_{u\in V(G_N)}d_G(u,v)$ to be the distance between a single vertex $v$ and a subgraph $G_N\subset G$. For a given underlying $G$ and a rumor graph $G_N$, we define the {\it rumor graph boundary} as follows.

\begin{definition}
Let $\mathcal{B}(G_N)=\{v\in V(G)\vert  d(v,G_N)=1\}$, and we call $\mathcal{B}(G_N)$ the rumor graph boundary of $G_N$. If a node $v\in \mathcal{B}(G_N)$ and the neighbor of $v$ in $G_N$ is not a leaf of $G_N$, then we call $v$ a {\it pseudo-leaf} of $G_N$. Lastly, we define $\bar{G}_N=G_N\cup \{ \text{all pseudo-leaves of } G_N\}$, and call $\bar{G}_N$ the {\it rumor closure} of $G_N$.
\end{definition}

Consider Fig. \ref{fig:treeEg} as an example, the rumor graph boundary is $\mathcal{B}(G_4)=\{v_0,v_4,v_6,v_7,v_8,v_9\}$, where $v_0$,$v_4$ are pseudo-leaves. The boundary of the rumor graph offers valuable perspectives on the dynamics of rumor dissemination, characterized by probabilistic properties that significantly influence the detection of the rumor source.

\subsection{Evaluation of Graph Boundary on Trees}

Within a tree graph, if we treat the rumor source as the root of the tree, then the rumor spreading is limited to movement from parent nodes to their child nodes, simplifying the evaluation of $P(G_N\,|\,v, T)$. This means that rumor propagation within a tree structure is limited to a single path, unlike in general graphs where multiple paths are possible. As illustrated in Fig. \ref{fig:treeEg}, assuming $v_1$ is the source of infection, according to this tree structure, $v_5$ will only be infected after $v_2$ has been infected. Furthermore, for being a leaf node, we must ensure that all child nodes of $v_5$, i.e., $v_8$ and $v_9$, are not infected by time $T$. Hence, we only need to consider the nodes on the boundary since they provide enough information about the rumor graph. 

\begin{figure*}[ht]
\centering
  \subcaptionbox{Given $K_{ij}$ and $d_i$, the variation of the value of \eqref{eq:leaf} against $T$.\label{fig:eqleaf_T}}[.42\linewidth]{\includegraphics[trim={0 0 0 0em},clip,width=0.8\linewidth]{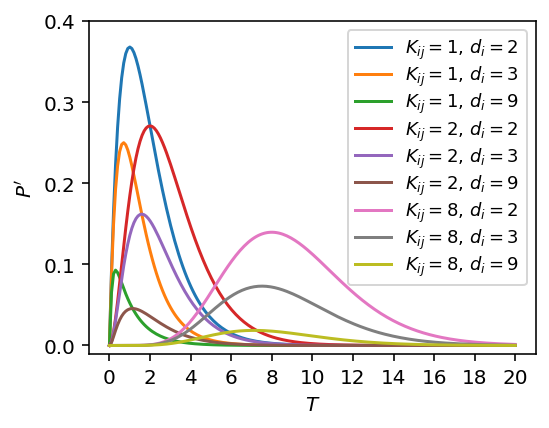}}\hfill
  \subcaptionbox{Given $K_{ij}$ and $T$, the variation of the value of \eqref{eq:leaf} against $d_i$. \label{fig:eqleaf_d}}[.42\linewidth]{\includegraphics[trim={0 0 0 0em},clip,width=0.8\linewidth]{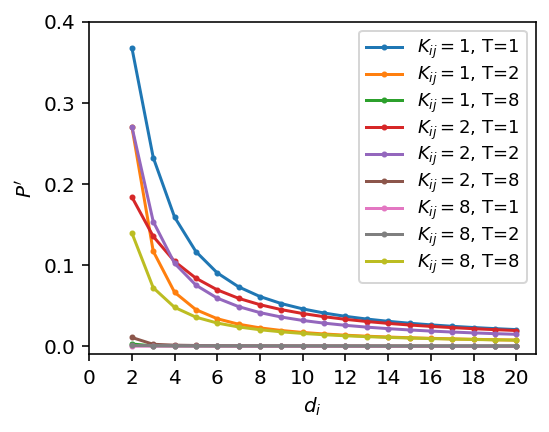}}
  \caption{Let $P'$ denote the numerical value of \eqref{eq:leaf} with corresponding $K_{i,j}$, $d_i$ and $T$. This figure shows the variation of $P'$.}
  \label{fig:eqleaf}
\end{figure*}

To simplify the formulation, we define $K_{ij}=d(v_i,v_j)$ to be the distance between $v_i$ and $v_j$. Hence, with the source node $v_j$ (that is, the root of the rumor tree $G_N$) and a specific time $T$, the probability that node $v_i$ act as a leaf in $G_N$ and that all its susceptible children $v_l\in \mathcal{B}(G_N)$ can be calculated by
\begin{equation}
\label{eq:leaf}
P(\textbf{v}_i\displaystyle\bigcap_{v_l\in\textrm{child}(v_i)} {\bar{\textbf{v}}}_l\,|\,\textbf{v}_j,T) 
= \int_0^T \frac{t^{K_{ij}-1}e^{-t}}{(K_{ij}-1)!}e^{-(T-t)(d_i-1)} \mathrm{d}t.
\end{equation}
To grasp the concept presented in \eqref{eq:leaf}, we re-label the shortest path from $v_j$ to $v_i$ as $v_1v_2v_3\ldots v_{K_{i,j}+1}$, where $v_1=v_j$ and $v_{K_{i,j}+1}=v_i$. Then, it is crucial to acknowledge that the spreading time from $v_j$ to $v_i$, represented by the sum  
$$
\tau_{1,2}+\tau_{2,3}+\dots+\tau_{K_{i,j},K_{i,j}+1}
$$ 
follows the Erlang distribution characterized by a shape parameter $K_{ij}$ and a rate parameter $\lambda=1$, under the condition that each individual transmission time $\tau_{1,2},\tau_{2,3},\dots+\tau_{K_{i,j},K_{i,j}+1}$ is exponentially distributed with the same parameter $\lambda$. The probability density function of this Erlang distribution for $\lambda=1$ is 
$$
\mathsf{Erlang}(K_{ij},\lambda=1)=\frac{t^{K_{ij}-1}e^{-t}}{(K_{ij}-1)!},
$$
i.e., the first term in the integration of (\ref{eq:leaf}). The mean of $\mathsf{Erlang}(K_{ij},\lambda)$ is $\frac{K_{i,j}}{\lambda}$, thus we can deduce that the value of (\ref{eq:leaf}) will reach to its maximum around $T\approx \frac{K_{i,j}}{\lambda}$ when $K_{i,j}$ and $d_i$ is fixed. For example, we can observe that for each pair of $(K_{i,j},d_i)$ in Fig. \ref{fig:eqleaf}, the value of the function reaches its maximum around $T\approx K_{i,j}$. This phenomenon can be understood as follows: Since each transmission time along an edge is $\tau$, it follows an exponential distribution with parameter $\lambda$. When there are $K_{i,j}$ edges with $\tau$ values in between, the total expected time would be $K_{i,j}\cdot \frac{1}{\lambda}$. Therefore, in practical situations, if the value of $\lambda$ changes, we can expect that the time at which the maximum value of (\ref{eq:leaf}) occurs will also change accordingly. However, the $K_{i,j}$ value represents the distance in the graph, which is derived from observed data and will not change as a result. The value of $K_{i,j}$ is determined by finding the shortest path between $v_j$ and $v_i$ in the graph and counting the number of edges in that path. 

The second term $e^{-(T-t)(d_i-1)}$, in the integration of (\ref{eq:leaf}), guarantees that all $d_i-1$ children of $v_i$ remain susceptible at time $T$. For example, we can set $K_{5,1}=2$, $d_5=2$ in (\ref{eq:leaf}) to compute the probability of $v_5$ being the leaf of $G_4$ in Fig. \ref{fig:treeEg}. On the other hand, if $v_i$ is a pseudo-leaf, we treat this case as equivalent to considering $\mathsf{parent}(v_i)$ as a degree-$2$ leaf. In this way, we can still use Equation (3) to calculate the probability that by time $T$, $v_i$ is observed as a pseudo-leaf. Specifically, we treat $\mathsf{parent}(v_i)$ as a leaf of $G_N$ and set its degree to $2$, then apply Equation (3) for the calculation.
The following lemma describes the properties of \eqref{eq:leaf} with respect to parameters $T$ and $d_i$.

\begin{lemma}\label{lemma:properties}
Given source $v_j$ and at time $T$, the probability of a node $v_i$ being a leaf node in the rumor graph $G_N$, but not a leaf in $G$, has the properties in terms of observation time $T$, the node depth in the tree $K_{ij}$ and the degree of the node $d_i$.
\begin{itemize}

\item For fixed $K_{ij}$ and $d_i$, Equation \eqref{eq:leaf} can be seen as a function of $T$. There exists a critical point $T_{\text{max}}$ such that the function increases for $ T < T_{\text{max}}$  and decreases for $T > T_{\text{max}}$. Moreover, the value of the function decreases to $0$ as $T$ approaches infinity.

\item For fixed $K_{ij}$ and $T$, Equation \eqref{eq:leaf} can be seen as a function of $d_i$, and the function is monotonically decreasing.
\end{itemize}
\end{lemma}
The formulation in \eqref{eq:leaf} suggests that the likelihood of the rumor reaching the leaf node is greater when there is sufficient time. Intuitively, extending $T$ provides the rumor more opportunity to propagate across successive edges within a specified distance $K_{ij}$. Conversely, a higher degree restricts the leaf node by increasing the number of susceptible child nodes it must maintain. Hence, as $T$ increases from $0$, the value of (\ref{eq:leaf}) also increases from $0$ until it reaches its maximum, i.e., the current time $T$ is ``adequate'' for the given $K_{i,j}$ which is around $K_{i,j}$ since we assume that $\lambda=1$. Then, the value of (\ref{eq:leaf}) starts decreasing to $0$ in the long run, which means the probability of spreading the rumor to a finite amount of nodes that use infinite time is almost equal to $0$. Hence, there will be no leaf node after a very long rumor-spreading time, meaning all nodes will be infected by then. 
The second property in Lemma~\ref{lemma:properties} can be understood similarly. We illustrate the two statements in Lemma~\ref{lemma:properties} by plotting the graph of \eqref{eq:leaf} in \figref{fig:eqleaf} with different tuples of $K_{ij}$, $d_i$, and $T$.

\captionsetup[subfigure]{skip=-0pt,position=bottom,belowskip=0pt}

Lastly, we introduce a simple but useful formulation for computing the likelihood of being the source between two adjacent nodes. Let $u,v$ denote two adjacent nodes in a connected graph $G_N$, and $(u,v)$ be a bridge in graph $G_N$. Then $G_N$ can be composed of two connected subgraphs plus the bridge $(u,v)$, i.e., $G_N=G^1_N\cup \{(u,v)\} \cup G^2_N$ where $u\in V(G^1_N)$ and $v \in V(G^2_N)$. We have

\begin{equation}
\label{eq:adj_prob_int}
    P(G_N | u,T)=P(G^1_N | u,T)\cdot \int_0^Te^{-x}P(G^2_N | v,T-t)\mathrm{d}t.
\end{equation}
The above equation provides a tool for comparing $P(G_N|v,T)$ and $P(G_N|u,T)$  between two adjacent nodes $u,v$.

\subsection{Maximum Likelihood Estimation for Trees}

In a specific scenario of $G_N$ where the source node $v$ serves as the root of the tree, the probability $P(G_N\,|\,v, T)$ can be determined solely based on the probabilities of all leaf nodes being infected by time $T$, as explained in the upcoming lemma.

\begin{lemma}\label{lemma:independen}
For any two leaves nodes $v_i$ and $v_j\in V(G_N)$, if the lowest common ancestor of $v_i$ and $v_j$ is the root $v$,  then $P(v_i \cap v_j \vert v, T)=P(v_i \vert v, T)\cdot P(v_j \vert v, T)$.
Otherwise, if $d(v_i,v)=d(v_j,v)$ and $T\leq d(v_i,v)-1$, then we have $P(v_i \cap v_j \vert v, T)\geq P(v_i \vert v, T)\cdot P(v_j \vert v, T)$.
\end{lemma}

The above lemma states that the infection of $v_i$ and $v_j$ are independent of one another when the lowest common ancestor of $v_i$ and $v_j$ is the root node. Note that when both $v_i$ and $v_j$ are pseudo-leaves of $G_N$, we can have the same conclusion as above. We introduce a class of graphs such that the likelihood of being the source of its centered node can be computed efficiently.

\begin{definition}
\label{def:star}
A connected simple graph $G$ is a {\it starlike graph} if $G$ satisfies \cite{starlike}:
\begin{itemize}
    \item $G$ has a spanning tree $\mathcal{T}$ such that $\mathcal{T}$ is homeomorphic to a star graph with the center node denoted as $v_c$.
    \item For each node $v\in V(G)$, $d_G(v,v_c)=d_{\mathcal{T}}(v,v_c)$.
\end{itemize}
In addition to general graph, a tree is said to be \textit{starlike} if and only if it has at most one node, say $v_c$, of degree three or more. Lastly, a \textit{star graph} is a specific type of starlike tree such that $v_c$ is directly connected to all other nodes, and none of the other nodes are connected to each other.
\end{definition}

If both $G$ and $G_N$ are starlike trees, and $G_N$ contains the centered node $v_c$ of $G$, then we can compute $P(v_c\vert G_N, T)$ by simply multiplying the probability (\ref{eq:leaf}) for all leaves. To simplify the notation, we denote the integrand in \eqref{eq:leaf} as $\hat{f}_i(v_j,t)$ for brevity. Note that the variables $K_{ij}$ and $d_i$ are not shown in $\hat{f}$ since the value of $K_{ij}$ and $d_i$ are fixed as $G_N$, $v_i$ and $v_j$ are given. By Lemma \ref{lemma:independen}, we have
\begin{equation}\label{eq:ml_tree}
\begin{array}{rl}
 P(G_N\,|\,v_c, T)
=\displaystyle\prod_{v_i\in\textrm{leaf}(G_N,v_c)} \left(\int_0^T\hat{f}_i(v_c,t)\mathrm{d}t\right),
\end{array}
\end{equation}
where $\textrm{leaf}(G_N,v_c)$ is a set containing all leaf nodes in $G_N$, when $v_c$ is considered to be the root of $G_N$. Since for each pair of leaves, their lowest common ancestor must be $v_c$. In the following, we present a message-passing technique aimed at calculating a solution that might not be as optimal as the one in (\ref{eq:ml}), but holds the potential to deliver the best possible global solution for certain graph topologies, such as line graphs or star-shaped graphs. This message-passing approach is motivated in part by maximizing the lower boundary mentioned in Lemma \ref{lemma:independen}. Subsequently, we introduce a heuristic for starlike graphs, which approximates either a given tree graph or a general graph as a starlike configuration and then employ the message-passing technique.

\subsection{Message-Passing Algorithm on Tree Graphs}

To identify the source within a tree $G_N$, it is essential to compute the $P(G_N|v,T)$ for all $v\in V(G_N)$. Under the assumption that nodes only obtain information from their immediate neighbors, we propose a message-passing algorithm that computes $P(G_N|v,T)$. The proposed algorithm includes two message-passing phases: top-down (from the root to leaves) and bottom-up (from leaves to the root).  Each node $v_i$ in the following algorithm has an attribute, i.e., the distance $k$ from the root $v_r$. Thus, $v_i.k$ indicates the distance from $v_i$ to the root node $v_r$. Initially, the attributes $k$ for all nodes are set to be $0$ based on our assumption.
The procedure is summarized in Algorithm~\ref{alg:passing1}.

\begin{algorithm}[!]
\caption{Message-Passing Algorithm for Likelihood of $v_r$ on Tree $\mathsf{MP}(v_r,\bar{G}_N,T)$ }\label{alg:passing1}
\begin{algorithmic}[1]
\STATE Let $v_r$ be the root and send a $M^{down}_{v_r\rightarrow \mathsf{child}(v)}=v_r.k+1$ to its child.
\STATE Whenever a node $u$ receive a message from its parent, $u$ send the message $M^{down}_{u\rightarrow \mathsf{child}(u)}=u.k+1$ to its child.
\STATE Repeat 2. until every non-root node in $\bar{G}_N$ received a message from their parent.

\STATE For each leaf  $u\in V(\bar{G}_N)$, set $M^{up}_{u\rightarrow \mathsf{parent}(u)}=\int_0^T\hat{f}_u(v_r,t)\mathrm{d}t$ and send the message $M^{up}_{u\rightarrow \mathsf{parent}(u)}$ to $\mathsf{parent}(u)$.

\STATE For each pseudo-leaf  $u\in\bar{G}_N$, set $M^{up}_{u\rightarrow \mathsf{parent}(u)}=\int_0^T\hat{f}_{parent(u)}(v_r,t,d_{parent(u)}=2)\mathrm{d}t$ and send the message $M^{up}_{u\rightarrow \mathsf{parent}(u)}$ to $\mathsf{parent}(u)$.

\STATE For each non-(pseudo)leaf and non-root node $u\in V(\bar{G}_N)$, if $u$ receive messages from all its children, then $u$ send the message $M^{up}_{u\rightarrow \mathsf{parent}(u)}=\prod_{w\in \mathsf{child}(u)} M^{up}_{w\rightarrow u}$ to its parent.

\STATE Return the value: $\prod_{w\in \mathsf{child}(v_r)} M^{up}_{w\rightarrow v_r}$
\end{algorithmic}
\end{algorithm}

During the initial top-down phase, which encompasses the first three steps in Algorithm \ref{alg:passing1}, the root of the tree sends messages indicating the distance from itself to the (pseudo)leaves of $\bar{G}_N$, enabling the calculation of $k$ for each node. Subsequently, in the bottom-up phase, each node collects the probabilities received from its children and passes this aggregated information to its parent. This process is repeated until the root has received information from all its children. Furthermore, the proposed Algorithm \ref{alg:passing1} can be implemented via a depth-first search (DFS) initiated at $v_r$ on $\bar{G}_N$.

\subsection{A Special Case and Properties}\label{sec:property}

This section is dedicated to evaluating the analytical performance of our ML estimator derived in \eqref{eq:ml_tree}. We commence by applying \eqref{eq:ml_tree} to a particular case: a 2-regular tree, essentially a line graph, demonstrating that our findings align with both intuitive expectations and existing literature \cite{Shah2011, pd_jstsp2022}, as corroborated by Theorem \ref{thm:like_ratio_reg}. Subsequently, we delve into the ML estimator's behavior in a line-shaped rumor graph, illustrating that the value $P(G_N|v_e,T)$ of a leaf node $v_e$ may either exceed or fall short of that of a centrally located node, depending on the underlying tree graph.

\begin{proposition}\label{prop:two}
If $G$ is a $2$-regular tree, and $G_N$ does not contain any leaf of $G$. Then, the estimated rumor source of $G_N$ is the node(s) in the middle of the line. However, if $G_N$ contains one leaf, say $v_e$ of $G$, then $v_e$ is the estimated rumor source as time $T$ goes to infinity. 
\end{proposition}

The above proposition can either be deduced from Theorem \ref{thm:like_ratio_reg} or directly computed by (\ref{eq:ml_tree}). Note that if one of $G_N$'s leaves is also a leaf of $G$, then the estimated rumor source may not be the node in the middle. The estimated source will be on the path from the leaf to the middle node of $G_N$, which depends on the time $T$ and demonstrates the ``boundary effect'' described in \cite{pd_ASONAM}.   

Next, we explore further the property of leaf nodes and pseudo-leaves. As established in Lemma~\ref{lemma:properties}, the value of \eqref{eq:leaf} first increases and then decreases with its distance to the root, which makes characterizing the ML estimator more complicated. Let us consider a simple example where $G$ is a general tree, and $G_N\subset G$ is a line graph with three nodes where $v_y$ is in the middle and $v_x$, $v_z$ are connected to $v_y$. Let $d_x$, $d_y$, and $d_z$ denote the degree of $v_x$, $v_y$, and $v_z$ respectively. If we treat $v_y$ as the root, then all neighbors of $v_y$, except $v_x$ and $v_z$, are pseudo-leaves. By considering all pseudo-leaves and leaves of $G_3$ we can have the closed-form expression of $P(G_3| v_i, T)$ for $i=x,y,z$ as follows: 
\begin{align*}
    &P(G_3| v_y,T)\\
    &=\frac{e^{-(d_x+d_y+d_z-4)T}}{(d_x-2)(d_z-2)}(e^{(d_x-2)T}-1)(e^{(d_z-2)}T-1),
\end{align*}
and 
\begin{align*}
    &P(G_3| v_z, T)\\
    &=\frac{e^{-d_zT}-e^{-(d_y+d_z-2)T}}{(d_x-2)(d_y-2)}-\frac{e^{-d_zT}-e^{-(d_x+d_y+d_z-4)T}}{(d_x-2)(d_x+d_y-4)}.
\end{align*}
Here we omit the close form of $P(G_N| v_x, T)$, since $v_x$ is symmetric to $v_z$ and $P(G_N| v_x, T)$ will be similar to that of $v_z$.
Observe that, the formulation reduce to (\ref{eq:d_reg_close_form}) when $d_x=d_y=d_z$, such as the constant $k=\frac{1}{(d_x-2)(d_y-2)}$ in $P(G_3| v_y, T)$. Thus, in the case of $d_x=d_y=d_z$, $v_y$ is the ML estimator of the source for the observed $G_3$. Otherwise, we can numerically show that if there is no further assumption on the degree of each node, then each of them is possible to be the ML estimator depending on their degrees. For example, when $d_x=d_z=3$ and $d_y=4$, we have $P(G_N| v_z, T)>P(G_N| v_y, T)$. The above example demonstrates that the number of pseudo-leaves significantly impacts the value of $P(G_N|v, T)$ for $v\in V(G_N)$. In the following, we combine the results in Lemma \ref{lem:T_inde}, \ref{lemma:independen} to compute $P(G_N|v,T)$ for each $v$ in a star graph.

\begin{theorem}\label{thm:central-is-source}
If $G$ is a $d$-regular tree graph and $G_N\subset G$ is a star graph. Let $v_c$ denote the centered node and $v_p$ be one of its neighbors. Then, we have 
\begin{align*}
 P(G_N| v_c, T)&= \frac{e^{-(N(d-2)+2)T}\cdot (e^{(d-2)T}-1)^{N-1}}{(d-2)^{N-1}},\\
 P(G_N| v_p, T)&= \frac{e^{-(N(d-2)+2)T}\cdot (e^{(d-2)T}-1)^{N-1}}{(N-1)(d-2)^{N-1}}. 
 \end{align*}
\end{theorem}
The above theorem directly leads to the result that $v_c$ is the ML estimator for the observed star graph $G_N$.

\section{Approximations and Asymptotics by Starlike Tree Graphs} 
\label{sec:starlike}

Starlike graphs, as described in Definition \ref{def:star}, creates a ``star" pattern, with the central node as the ``hub" and the other nodes as the ``spokes". In the context of online social networks like web blogs, e.g., \cite{blogstar}), and X (formerly Twitter), e.g., \cite{twitterstar,virusstarlike0,virusstarlike1,cascadestar,fraudgraph}), starlike graphs have also been observed to arise whereby a minority group of users (the central nodes) each has a large number of followers (the other nodes). 

X (formerly Twitter) can be modeled as a social network graph, where the users are represented as nodes, and the interactions (such as following, retweeting, and replying) between them are represented as edges. This user may be an influencer or high-profile individual whose tweets attract a large following. This is known as a ``scale-free" network. For example, on Twitter, the majority of users have relatively smaller follower counts as compared to the central users, and they follow the central users, thus forming the leaves of the starlike graph (so-called {Superstars})  \cite{twitterstar,virusstarlike0,virusstarlike1}. Thus, X (formerly Twitter) is an illustrative example of a network with starlike graphs where influential users have numerous followers who form the network's periphery.

The presence of these starlike structures suggests that individual influencers can have a significant impact on the cascade of information in a large online social network \cite{twitterstar,virusstarlike0,virusstarlike1,cascadestar,fraudgraph}. If a hub is an early adopter or influencer, its actions can have a disproportionate effect on the spread of information. For instance, a popular blogger sharing a new article can trigger a swift cascade of information among their numerous followers. The following result gives a necessary and sufficient condition to topologically characterize any starlike graph that we leverage to devise an algorithm for solving (\ref{eq:ml}) approximately.
\begin{theorem}[Corollary in \cite{starlike}]
\label{thm:starlike}
A tree is starlike if and only if it has at most one vertex of degree three or more.  
\end{theorem}
A starlike tree is obtained by attaching linear graphs (also called arms \cite{Srikant_ICASSP2016}) to this central vertex. We present a starlike approximation algorithm to transform a general graph $\bar{G}_N$ into a starlike tree $G^{\star}$ to approximate $P(\bar{G}_N | v, T)$ of $v\in V(G_N)$. 

\begin{algorithm}[!]
\caption{Starlike Tree Approximation for $P(G_N|v_r,T)$ on General Graphs}
\label{alg:star_contract}
\begin{algorithmic}[1]
\REQUIRE $\bar{G}_N$, $v_r$, $T$
\STATE $\mathsf{BFS}(\bar{G}_N,v_r)$
\STATE Let $G^{\star}=\{v^{\star}_r\}$
\STATE $v^{\star}_r.\text{infected}=\textbf{TRUE}$
\FOR{(pseudo)leaf $u\in V(\bar{G}_N)$}
    \STATE Let $p_{v_r^{\star},u^{\star}}$ be a path from $v_r^{\star}$ to $u^{\star}$ with length $d_{\bar{G}_N}(v_r,u)$ 
    \IF{$u$ is a leaf in $\mathsf{BFS}(\bar{G}_N,v_r)$}
        \STATE Set all nodes on $p_{v_r^{\star},u^{\star}}$ to be infected
        \STATE $G^{\star}=G^{\star}\cup p_{v_r^{\star},u^{\star}}$
        \STATE Set $d_G^{\star}(u^{\star})=d_{\Bar{G}_N(u)}$
    \ELSE
        \STATE Set all nodes on $p_{v_r^{\star},u^{\star}}$ to be infected except $u^{\star}$
    \ENDIF
\ENDFOR
\STATE $\Tilde{P}(G_{\text{N}}|v, T)=\mathsf{MP}(v^{\star}_r,G^{\star},T)$
\end{algorithmic}
\end{algorithm}
In Algorithm \ref{alg:star_contract}, we first apply the $\mathsf{BFS}$ graph traversal starting from $v_r$ on $\bar{G}_N$ to compute the distance $d_{\bar{G}_N}(v_r,u)$ from $v_r$ to $u$ for all $u\in V(\bar{G}_N)$. In the \textbf{for}-loop starting from line 4, we construct a new tree graph $G^{\star}$ based on the distance $d_{\bar{G}_N}(v_r,u)$ and the status (infected or not) for each (pseudo)leaf. We denote $v_r^{\star}$ and $u^{\star}\in V(G^{\star})$ as the corresponding nodes of $v_r$ and $u\in V(\bar{G}_N)$. Since the \textbf{for}-loop only considers (pseudo) leaf $u$ of $\mathsf{BFS}(\bar{G}_N)$, each corresponding $u$ only been considered once in line 4. Hence, the degree of $u^{\star}\in V(G^{\star})$ is at most 1 and all other nodes on the path $p_{v_r^{\star},u^{\star}}$, except $v_r^{\star}$ and $u^{\star}$, have degree at most 2. Lastly, we can observe that $v^{\star}_r$ is the only node in $G^{\star}$ with a degree possibly greater or equal to $3$ since we only attach new nodes to $v^{\star}_r$. By Theorem \ref{thm:starlike}, we can conclude that $G^{\star}$ is a starlike tree. Lastly, we directly apply Algorithm \ref{alg:passing1} to $G^{\star}$ since it is a tree. We denote $\Tilde{P}(G_{\text{N}}|v, T)$ as the starlike tree approximation of the likelihood $P(G_{\text{N}}|v, T)$. 

For a given $v_r$, $G_N$, and $T$, we can analyze the complexity of Algo. \ref{alg:star_contract} as follows. The construction of the starlike tree involves a $\mathsf{BFS}$ traversal on $\bar{G}_N$ and creates a duplicate graph which cost $O(|V(\bar{G}_N)|+|E(\bar{G}_N)|)$. The last step of Algo. \ref{alg:star_contract} contains at most $|v(\bar{G}_N)|$ single-variable numerical integrations. The computational complexity of Line 14 is  $O(|E(\bar{G}_N)|\cdot T_{quad})$, where $T_{quad}$ represents the average computational cost of a single integration using $\mathsf{scipy.integrate.quad}$. Hence, the overall time complexity is $O(|E(\bar{G}_N)|\cdot T_{quad})$. The notation $T_{quad}$ indicates that we use adaptive quadrature methods to compute numerical integration. The methods used here and the precision can be customized as needed, but this part is beyond the scope of this paper, so we will not elaborate further here.



\begin{figure*}
\centering
   {\includegraphics[scale =0.85] {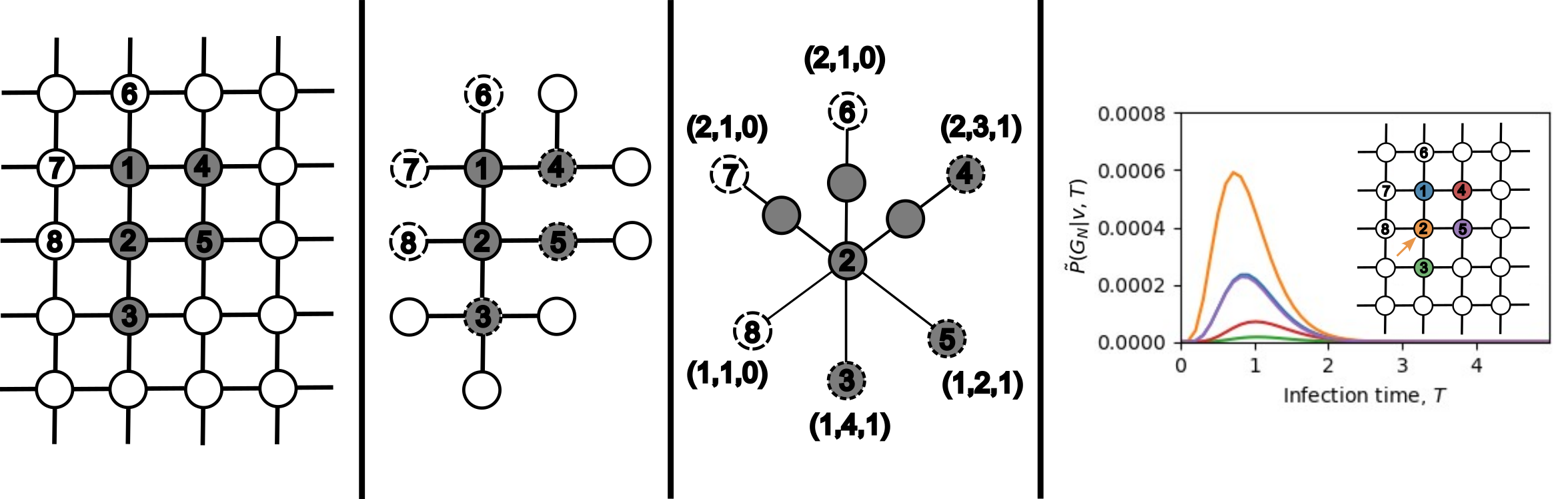}}
\caption{An example of how Algorithm 2 works on a grid graph. We first apply the BFS graph traversal starting from the root $v_2$ to obtain the rumor graph and its boundary. Then, we construct the starlike tree based on the resulting BFS tree. The tuple $(k,d, 0/1)$ beside each node, say $v$, represents $d_{G_N}(v,root)=k$, $deg(v)=d$ and whether $v$ is infected or not (1/0). The likelihood of being the source for each node is computed against time $T$. The estimated source is the node with the maximum value of $P(G_N|v,T)$. In this example, node $2$, which is indicated by an arrow, is the estimated source. We can also observe that the curve for $\tilde{P}(G_N|1,T)$ is almost overlapping with that for $\tilde{P}(G_N|v_5,T)$ due to the symmetric structure of $G_N$.}
\label{fig:grid_ex}
\end{figure*}

\subsection{Asymptotics in Performance of Likelihood Estimation}\label{sec:gammalogcvx}

\begin{figure}
\centering
   {\includegraphics[scale =0.29] {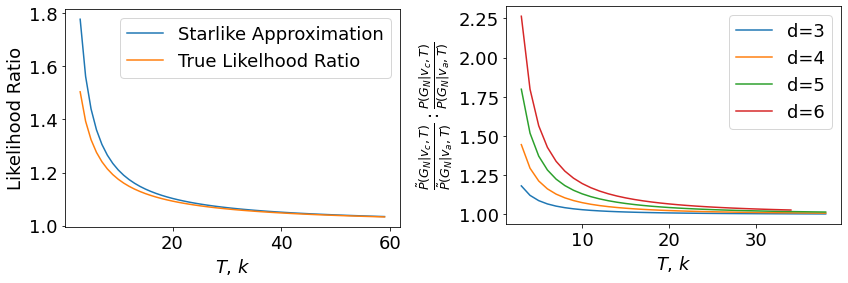}}
\caption{The left figure is the likelihood ratios computed by the starlike approximation and the true likelihood ratio when $d=3$ and $T=k$. Where as the right one illustrate the ratio $\frac{\tilde{P}(G_N|v_c,T)}{\tilde{P}(G_N|v_a,T)}:\frac{P(G_N|v_c,T)}{P(G_N|v_a,T)}$, for $d=3,4,5,6$. }
\label{fig:star_vs_true_ratio_big_d3}
\end{figure}

We evaluate the asymptotic behavior of the ratio $\frac{P(G_N|v_c,T)}{P(G_N|v_a,T)}$ between two nodes $v_c$ and $v_a$ in a starlike tree $G_N$, where $v_c$ is the center of the starlike tree, and $v_a$ is a neighbor of $v_c$. We compare the true ratio $\frac{P(G_N|v_c,T)}{P(G_N|v_a,T)}$ to the approximated ratio $\frac{\Tilde{P}(G_N|v_c,T)}{\Tilde{P}(G_N|v_a,T)}$ obtained by the starlike tree approximation. To simplify the analysis, we consider the starlike tree with $d$ arms, and each arm contains $k$ nodes, i.e., $N=kd+1$.  The approximated ratio can be computed by
\begin{align*}
    \frac{\Tilde{P}(G_N|v_c,T)}{\Tilde{P}(G_N|v_a,T)}&=\frac{(\frac{e^{-T}T^{k}}{k!})^d}{(\frac{e^{-T}T^{k+1}}{(k+1)!})^{d-1}\cdot \frac{e^{-T}T^{k-1}}{(k-1)!}} \\
    &= (\frac{k+1}{T})^{d-2}\cdot \frac{k+1}{k}.
\end{align*}
To evaluate the asymptotic behavior of $\frac{\Tilde{P}(G_N|v_c,T)}{\Tilde{P}(G_N|v_a,T)}$ we consider the scenario as $T$ goes to infinity. Then we have 
\begin{equation*}
\lim\limits_{T\rightarrow \infty}\frac{\Tilde{P}(G_N|v_c,T)}{\Tilde{P}(G_N|v_a,T)}= 1,   
\end{equation*}
since $\lim\limits_{T\rightarrow \infty} \frac{k}{T}=1$ as $\lambda=1$ in our assumption. 
Now, we consider the true ratio of the likelihood being the source between $v_c$ and $v_a$ in the following. For the center node $v_c$, we have $P(G_N|v_c,T)=\Tilde{P}(G_N|v_c,T)$. For $v_a$, we can leverage (\ref{eq:adj_prob_int}) to compute $P(G_N|v_a,T)$ as follows:

\begin{align*}
&P(G_N|v_a,T)= \frac{e^{-T}T^{k-1}}{(k-1)!} \int_0^Te^{-x} [\frac{e^{-(T-x)}(T-x)^k}{k!}]^{d-1}  \mathrm{d}t \\
&=k\cdot \left(\frac{e^{-T}}{k!}\right)^d \frac{e^{T(d-2)}T^{k-1}}{(d-2)^{k(d-1)+1}}\gamma(k(d-1)+1,(d-2)T),
\end{align*}
where $\gamma$ is the lower incomplete gamma function. The value of $\frac{P(G_N|v_c,T)}{P(G_N|v_a,T)}$ can be computed as 

\begin{align*}
  \frac{P(G_N|v_c,T)}{P(G_N|v_a,T)}&=\frac{[T(d-2)]^{k(d-1)+1}}{ke^{T(d-2)}\gamma(k(d-1)+1,(d-2)T)}\\
  &\approx \frac{T(-1)^{T(d-1)}\sum\limits_{i=0}^{\infty}(\frac{d-1}{-T})^{k+1}b_i(\eta)-1}{d-2}.
\end{align*}
for large $T$, $k$, where $\eta = \frac{d-2}{d-1}$, and $b_0(\eta)=1$ ,$b_1(\eta)=\eta$, $b_i(\eta)=\eta(1-\eta)b'_{i-1}(\eta)+\eta(2i-1)b_{i-1}(\eta)$ \cite{Temme1994}. 
The above result can be further computed by using series expansion or other asymptotic methods \cite{gammaasymptotic}, which is beyond the scope of the present paper. We evaluate the above two likelihood ratios numerically for the case $d=3$ in the left subgraph of Fig. \ref{fig:star_vs_true_ratio_big_d3}. The right subgraph shows the tendencies of the ratio $\frac{\tilde{P}(G_N|v_c,T)}{\tilde{P}(G_N|v_a,T)}:\frac{P(G_N|v_c,T)}{P(G_N|v_a,T)}$ are similar for different value of $d$. 

\section{Experiments}\label{sec:sim}

This section first provides numerical examples to illustrate the properties of the proposed estimator on starlike trees; then we compare the starlike tree approximation with existing methods in different synthetic and real-world networks.

\subsection{Numerical Results on Starlike Tree Graphs}

To demonstrate the capability of simplification of the starlike tree approximation in Algo. \ref{alg:star_contract}, we first list out the closed form of the estimator $\Tilde{P}(G_{\text{N}}|v, T)$ and the likelihood $P(G_{\text{N}}|v, T)$ for different starlike trees in Tab. \ref{tab:closeform_line_end}. We then plot the estimator $\Tilde{P}(G_{\text{N}}|v, T)$ numerically in Fig. \ref{fig:line-graph-leaf-infected} and \ref{fig:ML_star_graph_B3}. From Tab. \ref{tab:closeform_line_end}, we can observe that when $G$ is a line graph (a starlike tree with two arms), we have $P(G_{\text{N}}|v, T)=\Tilde{P}(G_{\text{N}}|v, T)$ for all $v\in G_N$.  If $G$ has more than two arms and $v$ is the central node of the starlike tree, then we still have $P(G_{\text{N}}|v, T)=\Tilde{P}(G_{\text{N}}|v, T)$. Otherwise, the approximation $\Tilde{P}(G_{\text{N}}|v, T)$ serves as a lower bound of $P(G_{\text{N}}|v, T)$ which justify the results in Lemma \ref{lemma:independen}.

\begin{table}[h]
\centering
\small\addtolength{\tabcolsep}{-5pt}

\begin{tabular}{|c|c|c|c|}
\hline
Fig. & $v$ & $\Tilde{P}(G_{\text{N}}|v, T)$ & $P(G_{\text{N}}|v, T)$ \\ \hline
\multirow{2}{*}{5(a)} & $v_0$ & $\frac{1}{2}T^2e^{-T}$   &  $\frac{1}{2}T^2e^{-T}$  \\ \cline{2-4} 
                      & $v_1$ & $T(1-e^{-T})e^{-T}$      &  $T(1-e^{-T})e^{-T}$ \\ \hline
\multirow{2}{*}{6(a)} &$v_0$  & $T^3e^{-3T}$             &  $T^3e^{-3T}$ \\ \cline{2-4} 
                      &  $v_1$  & $\frac{1}{4}T^4e^{-3T}$ & $(2e^T-T^2-2T-2)e^{-3T}$ \\ \hline

\end{tabular}
    \caption{Comparison between the true likelihood $P(G_{\text{N}}|v, T)$ and its approximation $\Tilde{P}(G_{\text{N}}|v, T)$, when considering starlike graphs shown in Fig. \ref{fig:line-graph-leaf-infected} and \ref{fig:ML_star_graph_B3}.}
    \label{tab:closeform_line_end}
\end{table}

Next, we consider the case of different starlike trees $G_N$ containing a leaf node of $G$. In Fig. \ref{fig:line-graph-leaf-infected} and \ref{fig:ML_star_graph_B3}, a node that is colored in white represents uninfected nodes. Otherwise, it is infected. In the scenario where exactly one leaf node of a line graph is infected, \figref{fig:line-graph-leaf-infected} presents the maximum likelihood probability versus spreading time for each node to be the source, as approximated by the starlike tree method. The node with the highest likelihood probability over time is identified as the detected source. 
In each line graph in ~\figref{fig:line-graph-leaf-infected}, node $0$ is an infected leaf node of the underlying network $G$, and node $i$ is connected to node $(i-1)$ for $i\geq 1$. If there are 1 or 2 infected nodes, the detected source is node $0$, the infected leaf node. However, when the number of infected nodes increases to 4 or 5, the detected source shifts to node $1$, which lies on the path from the leaf node to the central node of the rumor graph. We can also observe the same phenomenon in Fig. \ref{fig:ML_star_graph_B3-311-111}.

{
\captionsetup[subfigure]{skip=-0pt,position=bottom,belowskip=0pt}
\begin{figure*}[ht]
\centering
  \subcaptionbox{Line graph with $3$ infected nodes.\label{fig:ML_line_graph_one_leaf_infected-3-01}}[.33\linewidth]{\includegraphics[trim={0 0 0 0em},clip,width=0.98\linewidth]{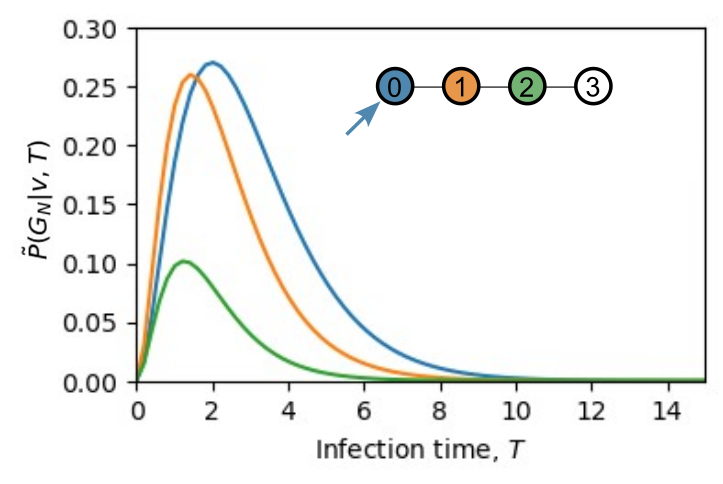}}\hfill
  \subcaptionbox{Line graph with $4$ infected nodes. \label{fig:ML_line_graph_one_leaf_infected-4-01}}[.33\linewidth]{\includegraphics[trim={0 0 0 0em},clip,width=0.98\linewidth]{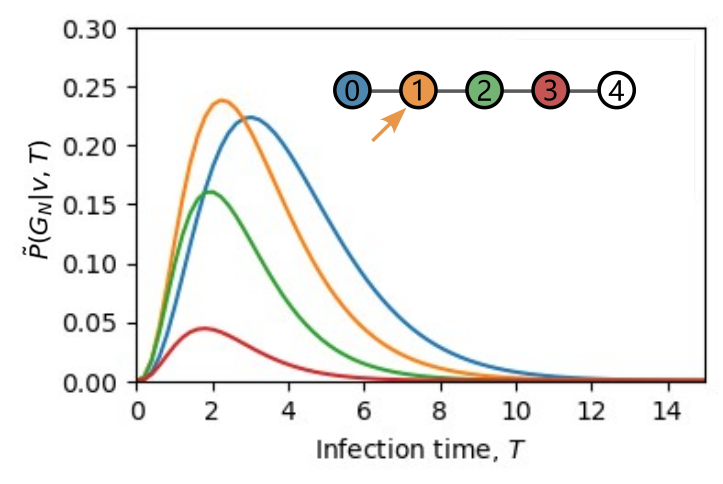}}\hfill
  \subcaptionbox{Line graph with $5$ infected nodes. \label{fig:ML_line_graph_one_leaf_infected-5-01}}[.33\linewidth]{\includegraphics[trim={0 0 0 0em},clip,width=0.98\linewidth]{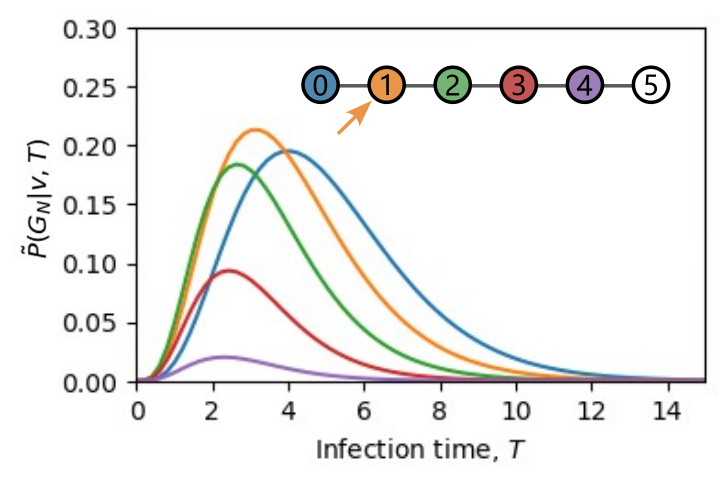}}\hfill
  \caption{$P(G_\text{N}|v, T)$ given by starlike tree approximation in line graphs where $G_\text{N}$ contains exactly one leaf of $G$. The estimated sources are indicated by arrows.}
  \label{fig:line-graph-leaf-infected}
\end{figure*}
}

\figref{fig:ML_star_graph_B3} provides an example of a starlike graph with three arms with difference lengths connected to the central node. These figures also include a curve representing $P(G_{\text{N}}|v, T)$ as a function of $T$ for a selected node $v$. In \figref{fig:ML_star_graph_B3-111-111}, the central node $0$ is identified as the estimated source as shown in Theorem~\ref{thm:central-is-source}.

In starlike tree graphs where $G_N$ does not contain any leaf node of $G$, the central node may not always be the estimated source. For example, \figref{fig:ML_star_graph_B3-311-111} is more unbalanced than \figref{fig:ML_star_graph_B3-211-111} regarding the size of the linear graphs attached to the central node. Consequently, the estimated source shifts from the central node $0$ to the non-central node $1$ in \figref{fig:ML_star_graph_B3-311-111}. Furthermore, in starlike tree graphs, when $G_N$ includes the leaf of $G$, the estimated source may shift due to the boundary effect~\cite{pd_ASONAM}. For example, the estimated source is node $4$ in \figref{fig:ML_star_graph_B3-311-011}, which differs from that in \figref{fig:ML_star_graph_B3-311-111}.

{
\captionsetup[subfigure]{skip=-0pt,position=bottom,belowskip=0pt}
\begin{figure}[ht]
\centering
  \subcaptionbox{\label{fig:ML_star_graph_B3-111-111}}[.47\linewidth]{\includegraphics[trim={0 0 0 0em},clip,width=1.0\linewidth]{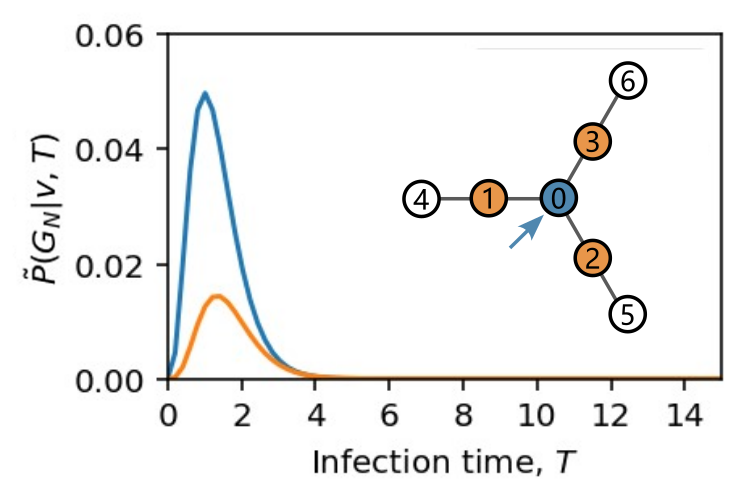}} \hfill
  \subcaptionbox{\label{fig:ML_star_graph_B3-211-111}}[.47\linewidth]{\includegraphics[trim={0 0 0 0em},clip,width=1.0\linewidth]{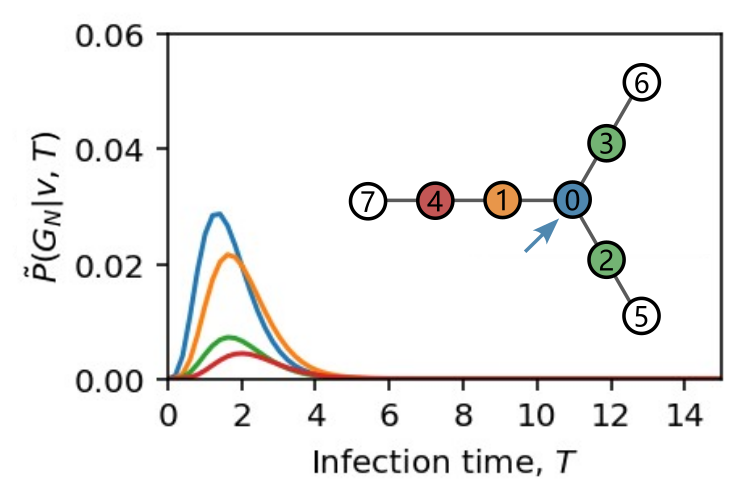}}\\
  \subcaptionbox{\label{fig:ML_star_graph_B3-311-111}}[.47\linewidth]{\includegraphics[trim={0 0 0 0em},clip,width=1.0\linewidth]{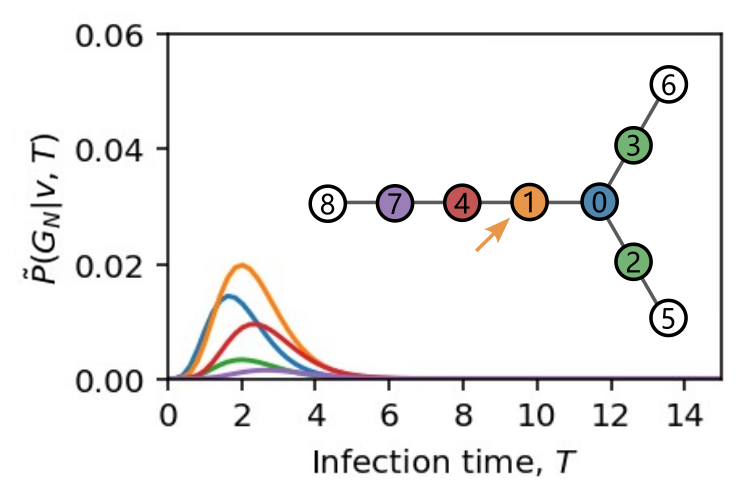}} \hfill
  \subcaptionbox{\label{fig:ML_star_graph_B3-311-011}}[.47\linewidth]{\includegraphics[trim={0 0 0 0em},clip,width=1.0\linewidth]{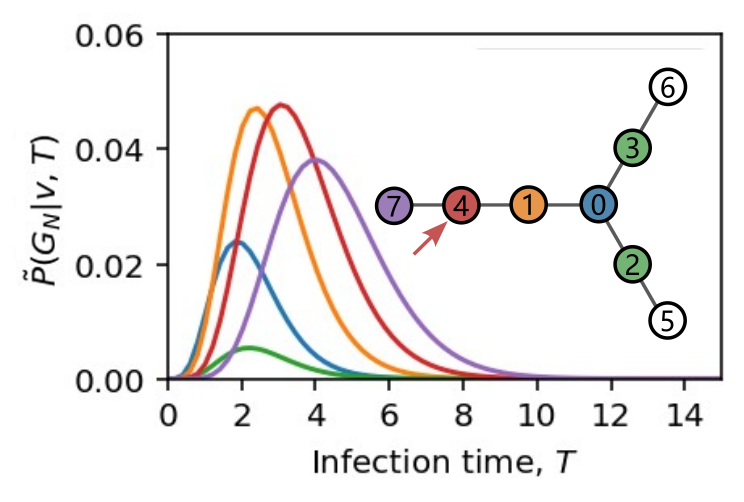}}
  \caption{$\Tilde{P}(G_\text{N}|v, T)$ in star/starlike tree graphs where the central node has 3 infected neighbors. The estimated sources are indicated by arrows.}
  \label{fig:ML_star_graph_B3}
\end{figure}
}

\subsection{Comparison with Existing Methods on Synthetic and Real World Networks}
To assess the performance of the starlike tree approximation and compare it with existing methods, we conducted simulation experiments on both synthetic networks and real-world networks. We adopt two different methods to simulate rumor spreading on networks, both of which follow the diffusion model described in Section II-A. The first method involves fixing the number of nodes in $G_N$, which represents the total number of infected nodes. In this approach, we uniformly select a node in $G$ as the source and begin the diffusion process until the number of nodes in $G_N$ reaches the desired count. We then record the topology of $G_N$, and the time $T$ taken, noting that in this case, the time $T$ is not fixed. The second method involves fixing the spreading time $T$, which is detailed in the subsection of real-world networks, resulting in $G_N$ of varying sizes. Due to space limitations, we fix $N$ for synthetic network simulations and $T$ for real-world network simulations.

To evaluate our proposed starlike tree approximation, we use two evaluation metrics. The first metric is the \textit{accuracy} (also known as the top-$1$ accuracy), which is the probability that the source estimator equals the true source. The second metric is the \textit{distance error}, which measures the graph distance between the source estimator and the true source. We choose the rumor centrality with $\mathsf{BFS}$ heuristic proposed in \cite{Shah2011,Shah2012}, the Jordan center-based approach proposed in \cite{zhu_jordan}, and the distance center as three methods for comparison with our approach. In instances of multiple nodes computed as the source, one is chosen at random to be the detected source.

\subsubsection{Synthetic Data}
\begin{figure*}[ht]
\centering
  \includegraphics[width=0.98\linewidth]{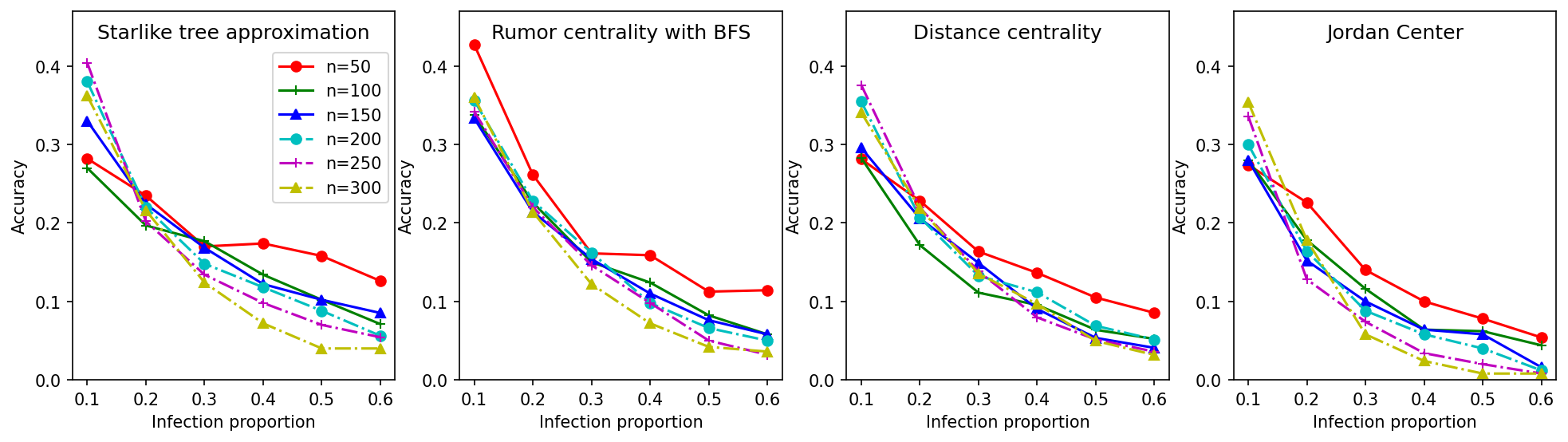}
    \caption{Accuracy of the rumor detection methods against the infection proportion in Erdős–Rényi model.}
    \label{fig:accuracy-infection_proportion}
\end{figure*}

For synthetic networks, we conducted simulation experiments on the Random Trees and Erdős–Rényi (ER) random graphs. Random Tree generates a tree selected uniformly from the set of all possible trees with the given number of nodes, while the latter model generate general graphs. In all synthetic networks, we generated underlying networks $G$ with varying numbers of nodes, specifically $n\in \{50, 100, 150, 200, 250, 300\}$. For ER random graphs, we set the probability of having edges between any pair of nodes to be $0.04$. Our analysis aims to evaluate the starlike tree approximation's accuracy against existing methods by examining various infection ratios, specifically $0.1, 0.2, 0.3, 0.4, 0.5$, and $0.6$. The accuracy of rumor source detection is depicted in \figref{fig:accuracy-infection_proportion}. Overall, as the infection proportion increases, the accuracy of all methods tends to decrease.

\begin{figure*}[ht]
\centering
  \includegraphics[scale =0.6]{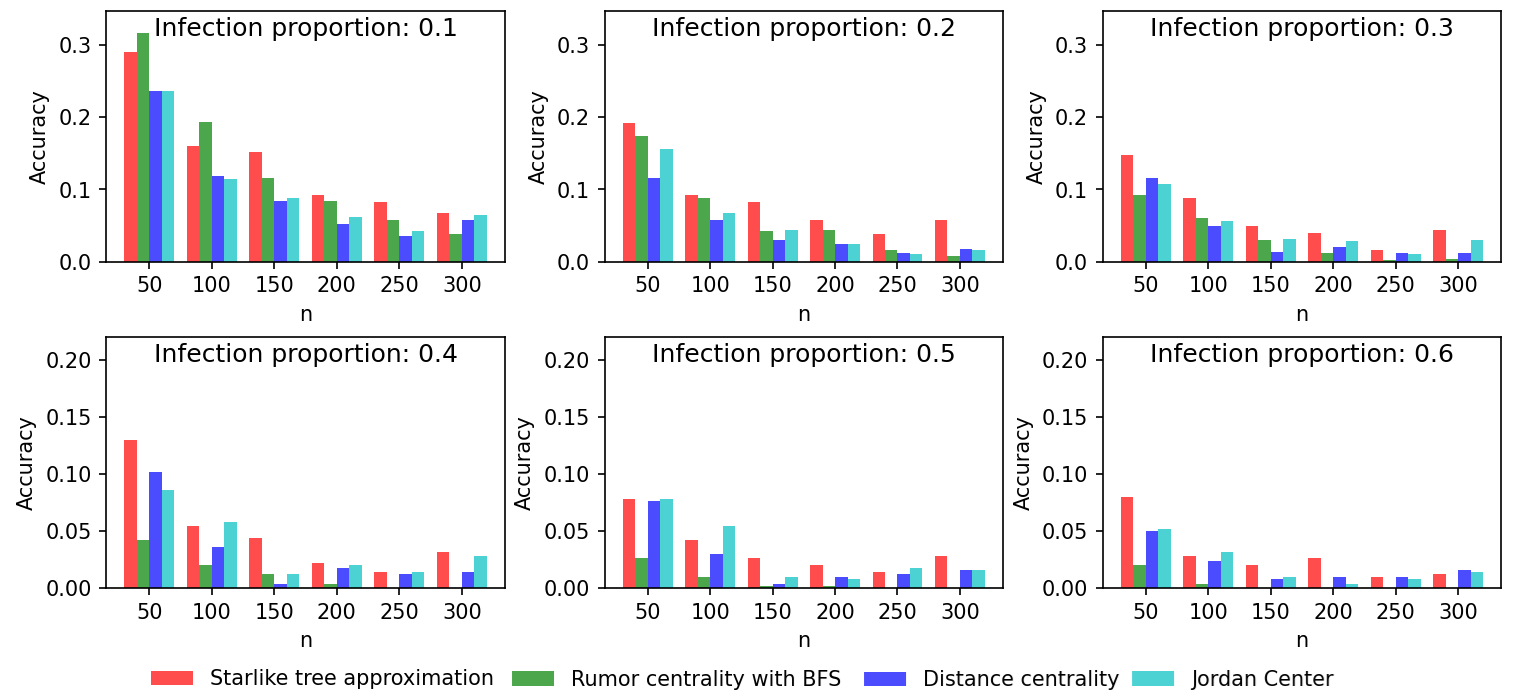}
    \caption{Accuracy of the rumor detection methods against the infection proportion in random tree graphs.}
    \label{fig:Accuracy_RT}
\end{figure*}

In the case of random trees, the starlike tree approximation demonstrates higher accuracy compared to other methods when the infection proportion ranges between 0.2 and 0.3, and the number of nodes varies from 50 to 300. This superiority can be attributed to the starlike tree approximation's utilization of information from the boundary of the rumor graph, its consideration of the phase-type distribution of spreading, and its resistance to the boundary effect, in contrast to the rumor centrality with the BFS heuristic method. Given these characteristics, it is not surprising that the starlike tree approximation outperforms other methods in the Random Tree scenario, particularly when the number of nodes and the infection proportion is around $0.3$ and $0.4$.

\begin{figure*}[ht]
\centering
  \includegraphics[scale =0.6]{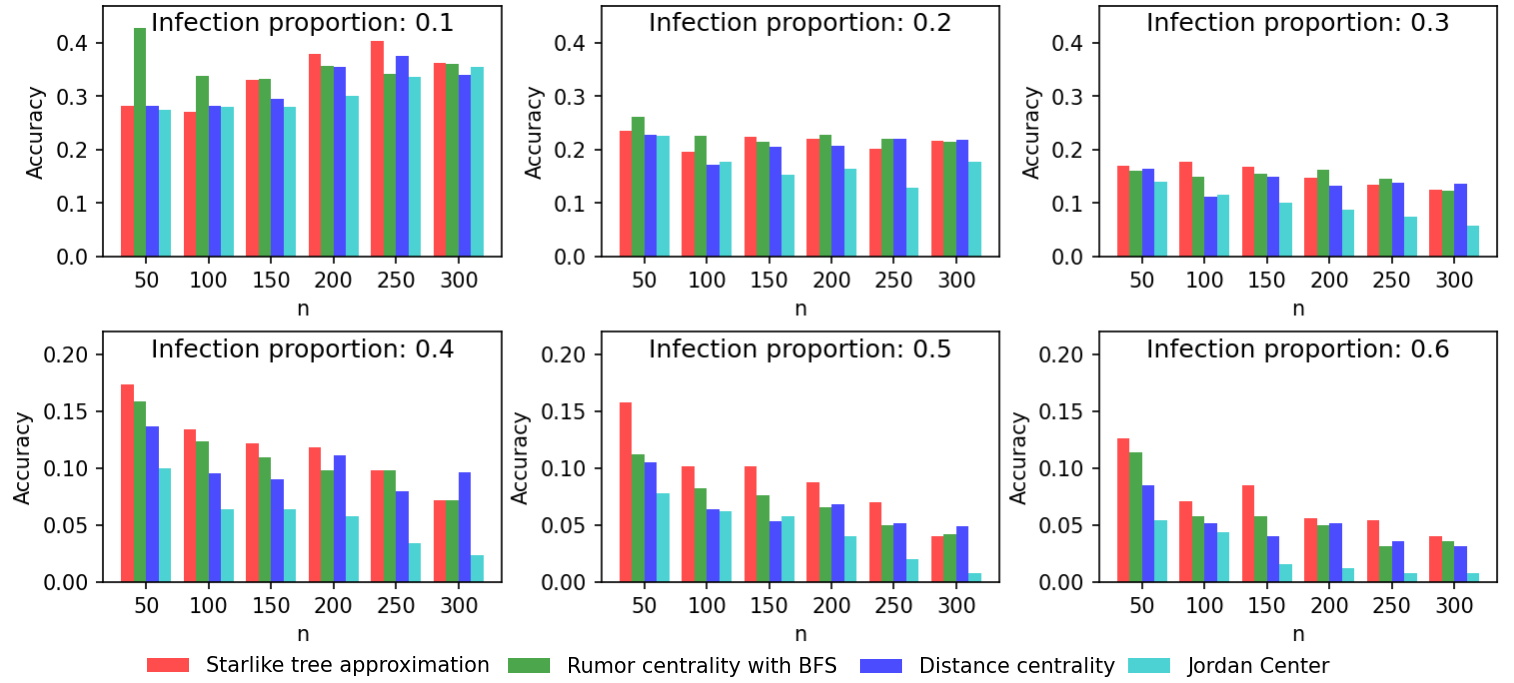}
    \caption{Accuracy of the rumor detection methods against the infection proportion in Erdős–Rényi model.}
    \label{fig:Accuracy_ER}
\end{figure*}

In the Erdős–Rényi model, as depicted in \figref{fig:Accuracy_ER}, the starlike tree approximation demonstrates comparable accuracy across various combinations of the number of nodes and infection proportion. Notably, in the Erdős–Rényi model, the starlike tree approximation exhibits improved accuracy when the infection proportion is 0.5, and the number of nodes in the underlying network ranges from 50 to 250.

\subsubsection{Real World Networks}
In the simulation experiments for rumor spreading and source detection on real-world networks, we tested three different online social network datasets: X (formerly Twitter)\cite{snap_face_twitter}, Facebook \cite{snap_face_twitter}, and Twitch gamer network \cite{twitch_dataset}. We ignore all attributes along with edges, such as edge weight or direction, in these datasets. Thus, we consider $G$ as an undirected simple graph. If the dataset contains more than one network (such as Twitch and X), we combine all the subnetworks together by relabeling the nodes to form $G$. The simulation process is described as follows: first, we are given an underlying network $G$ and a fixed spreading time $T$. At time $t = 0$, we uniformly select a node in $G$ as the source and simulate the spread of the rumor in the network. 
The spreading follows the SI model described in Section II-A. When the time reaches $t = T$, we stop the process and record all the nodes that have received the rumor, forming an induced subgraph of $G$, which we denote as $G_N$. Since the spreading process is a stochastic model, the number of nodes in $G_N$ varies in each simulation. Lastly, we apply different source estimation approaches described in \cite{Shah2011,Shah2012,zhu_jordan}  to find the source estimator given $G$, $G_N$, and $T$. For each given time $T$, we simulate the process $500$ times, recording the number of nodes and edges in $G_N$ each time, as well as the distance on $G$ between the estimators produced by the four methods and the true source. Finally, we calculate the average of each value and record them in the table.

We summarize the information of all three underlying networks: Facebook, Twitch, and X (formerly Twitter) and the simulation results of rumors spreading over three networks in Table \ref{tab:all_simu_results} below. We abbreviate the four methods as $\mathsf{STA}$, $\mathsf{BFS}$-$\mathsf{RC}$, $\mathsf{JC}$, and $\mathsf{DC}$, and compare them using the distance error.

\begin{table*}
\caption{Average error (in terms of number of hops) comparing Starlike Tree Approximation ($\mathsf{STA}$) to BFS Rumor Centrality (BFS-$\mathsf{RC}$), Jordan Center ($\mathsf{JC}$) and Distance Centrality ($\mathsf{DC}$) in different social networks. In the column of Ave. $\vert V(G_N) \vert$, the first number represents the average number of nodes in $G_N$, while the number in parentheses indicates the range of nodes in $G_N$ (as the number of nodes is not fixed).
}

\centering
\begin{tabular}{c|c|c|c|c|c|c|c|c|c}
 Network &$\vert V(G)\vert$ & $\vert E(G)\vert$ & $T$   & Ave. $ \vert V(G_N) \vert$ & Ave. $ \vert E(G_N) \vert$ & $\mathsf{STA}$ & $\mathsf{BFS}$-$\mathsf{RC}$ & $\mathsf{JC}$ & $\mathsf{DC}$
\\
\hline
\multirow{3}{*}{Facebook} & \multirow{3}{*}{4,039} & \multirow{3}{*}{88,234} & 1 & 109 (1\textasciitilde228) & 2,375   & \textbf{1.20} & 1.23 & 1.33 & 1.22 
\\
\cline{4-10}
        &  &  & 2 & 213 (52\textasciitilde372) & 6,518   & \textbf{1.24} & 1.34 & 1.40 & 1.27 
\\
\cline{4-10}
         & &  & 3 & 264 (27\textasciitilde489) & 7,829   & \textbf{1.26} & 1.36 & 1.54 & 1.32 
\\
\hline
\multirow{3}{*}{Twitch} & \multirow{3}{*}{34,118} & \multirow{3}{*}{429,113} & 1  & 85 (1\textasciitilde148) & 297   & 1.8 & 1.87 & 2.02 & \textbf{1.76} 
\\
\cline{4-10}
        &  &  & 1.5 & 157 (30\textasciitilde213)  & 771   & \textbf{1.84} & 2.22 & 2.18 & 1.90 
\\
\cline{4-10}
         & &  & 2 & 200 (18\textasciitilde268) & 1145   & \textbf{1.85} & 2.22 & 2.24 & 1.89 
\\
\hline
\multirow{3}{*}{X} & \multirow{3}{*}{81,306} & \multirow{3}{*}{1,768,149} & 1 & 94 (24\textasciitilde117) & 1,819   & \textbf{1.27} & 1.28 & 1.748 & 1.272 
\\
\cline{4-10}
        &  &  & 2 & 145 (29\textasciitilde164) & 3,703   & \textbf{1.256} & 1.26 & 1.494 & \textbf{1.256}
\\
\cline{4-10}
         & &  & 3 & 176 (140\textasciitilde190) & 4,809   & \textbf{1.33} & \textbf{1.33} & 1.59 & \textbf{1.33} 
\end{tabular}
\label{tab:all_simu_results}
\end{table*}

Across all networks and time steps, the $\mathsf{STA}$ method consistently achieves the lowest distance errors, indicating superior performance in source estimation. On the Facebook network, $\mathsf{STA}$ maintains the lowest errors with slight increases as time $T$ progresses, suggesting robustness even as the rumor spreads. In the Twitch network, while DC outperforms $\mathsf{STA}$ at $T=1$, $\mathsf{STA}$ shows better accuracy at later times $T=1.5$ and $T=2$. For the X network, $\mathsf{STA}$, $\mathsf{DC}$, and $\mathsf{BFS}$-$\mathsf{RC}$ demonstrate comparable performance with minimal error differences, whereas $\mathsf{JC}$ consistently exhibits higher errors across all networks.

In conclusion, the $\mathsf{STA}$ method emerges as the most reliable for accurate rumor source detection across different network types and sizes. The consistently low errors suggest that $\mathsf{STA}$ is effective regardless of the characteristics of networks, making it a valuable tool for rumor containment strategies. The source code of the above simulation can be publicly accessed at \url{https://github.com/convexsoft/Rumor-Source-Detection}

\section{Further Discussions}\label{sec:discussions}

\subsection{Limitations of the Proposed Method}

Despite the strengths of the starlike tree approximation method, several limitations are worth noting. First, this approach relies heavily on numerical integration, which can introduce computational challenges. As mentioned in Lemma \ref{lemma:properties}, the numerical integration value in Equation (\ref{eq:leaf}) becomes very small when either $d_i$ or $|T-K_{i,j}|$ grows large, leading to truncation errors. This sensitivity might affect the stability of the computation and result in inaccurate estimations.

Another limitation arises from the need for boundary information of the rumor graph $G_N$. This boundary information implies that the number of nodes required for computation slightly exceeds the actual number of nodes in $G_N$, leading to additional computational complexity. As a result, the method can be computationally expensive, particularly for large graphs. To address these potential limitations, we have incorporated a time complexity analysis of Algorithm \ref{alg:star_contract}, particularly focusing on the number of edges, which we believe will provide a clearer understanding of the computational efficiency of the proposed method. Lastly, our study is limited to the SI spreading model, in which the spread time between nodes follows an exponential distribution. Although this assumption is widely used in epidemic and rumor-spreading models, it may restrict the general applicability of our method to other types of spreading dynamics that deviate from this distribution.

\subsection{Extensions to Markovian Models and Network Applications}
The preceding sections outline the basic contagion source detection algorithm for the exponential spreading model, allowing for diverse variations without requiring significant modifications to the algorithm or its analysis. Examining various Markovian models for spreading rumors in networks is crucial for comprehending information dissemination, social dynamics, and communication patterns. In addition to the exponential distribution discussed in this paper, two Markovian models suitable for analysis in this context are hyper-Erlang distributions and phase-type distributions, traditionally employed in queueing theory \cite{kobayashi1, kobayashi2, networkmodel0}. Hyper-Erlang distributions are a class of probability distributions that generalize the traditional Erlang distribution by allowing for more flexible shapes. In the context of rumor spreading, hyper-Erlang distributions can be used to model the inter-event times between consecutive instances of rumor transmission. This approach considers the varying speeds at which rumors propagate through different graph edges. By fitting hyper-Erlang distributions to empirical data on rumor propagation times, we can determine additional topological features that accelerate or slow down the spread of rumors in networks.

 
Additionally, the probabilistic inference framework in this paper can be applied to other aspects of networking, including identifying the most crucial node for efficient data transmission. The progression from the pivotal node to the terminal node within the infection graph can be likened to {\it a series of connected exponential server queues.} As a result, we can study {\it the selection of the source node in the network to transmit data efficiently to a specified set of terminal nodes in the shortest possible time, considering exponential server timing and unique service rates for each server.} In particular, with a fixed aggregate service rate, the work in \cite{ordering1,ordering2,ordering3,ordering4} shows that the data departure process achieves its maximum stochastic speed when all the servers in the network are homogeneous (i.e., identical Markovian model at each node). It will be interesting to combine our work with that in \cite{ordering1,ordering2,ordering3,ordering4} to explore new networking applications.

Other interesting applications of the results in this paper involve correlated exponential random variables in a network, such as the Marshall-Olkin (MO) Copula Model for dependent failures and reliability assessment in networks in \cite{marshallolkin0,marshallolkin1,marshallolkin2}. Hence, identifying the origin of information spreading in networks is similar to the problem of pinpointing failure points in a network \cite{pd_JSTSP2018}. For example, it is possible to use Lemma \ref{lem:T_inde} and Lemma \ref{lemma:properties} in this paper with the MO Copula Model \cite{marshallolkin0} to better understand the dependencies between different graph components and their susceptibility to infection or failures originating from specific sources. It will be interesting to utilize a forward engineering methodology for network immunization problem, as described in \cite{pd_JSTSP2018, tanyusurvey, DAVA, MCWDST}, along with incorporating statistical correlations within a network, to enhance the accuracy of predictions and facilitate better decision-making in designing resilient and robust networks, especially in crucial applications where cascading failures could yield substantial consequences.

\section{Conclusion}\label{sec:conclusion}

This work presents a comprehensive probabilistic analysis of rumor spreading in networks, with a specific focus on the boundary behavior of rumor graphs. We develop a maximum likelihood estimation approach for rumor source detection, revealing that the likelihood ratio between nodes is time-independent in degree-regular tree networks. We introduced a distributed message-passing algorithm that is proven globally optimal for starlike trees. In addition to its optimal performance in starlike trees, the algorithm demonstrates effectiveness in general tree networks, showcasing its adaptability.

To extend the applicability of our analysis on tree networks, we proposed a starlike tree approximation that enables effective source detection in general networks, including those containing cycles. This extension allows us to handle more complex network topologies that go beyond simple tree structures. Moreover, by utilizing the asymptotic properties of the gamma function, we were able to analyze key graph-theoretic features in large-scale networks, aiding in the approximation of essential combinatorial aspects necessary for source estimation. Our extensive evaluations, conducted across a variety of graph structures, including those with intricate cycles and boundaries, confirm the robust performance of the starlike tree approximation. These results underscore the practical utility of our approach in real-world scenarios, demonstrating its capacity to perform reliably in diverse and complex networks.

\noindent
\textbf{Acknowledgements.} We would like to express our gratitude to Liang Zheng for her valuable insights and contributions to the problem formulation of this work. We also extend our thanks to Chao Zhao for his assistance in the experimental simulations presented in this paper.

\appendix

We sketch the proof of Lemma \ref{lem:T_inde} as follows. We first prove that for each leaf node $v\in V(G_N)$, the value of $P(G_N|T,v)$ is of the form $k\cdot e^{-(N(d-2)+2)T}\cdot (e^{(d-2)T}-1)^{N-1}$, for some scalar $k$. Then, we extend the result to the non-leaf node by decomposing the original graph into several small graphs, as shown in Fig. \ref{fig:lem_1_leaf}.

We prove the first statement by induction on the number $N$. For the base case $N=2$, we assume that $G_N$ contains two infected nodes $\{u,v\}$, connected by an edge $e_{u,v}=a$. We denote $2(d-2)$ edges, connecting to $v$ and $u$, other than $e_{u,v}$ by $x_i$ and $y_i$ for $i=1,2,\ldots d-1$ respectively. Without loss of generality, we have
\begin{align*}
    P(G_2|T,v)=\frac{1}{d-2}e^{-(2(d-2)+2)T}(e^{(d-2)T-1})^{2-1},
\end{align*}
where we can set $k=1/(d-2)$.
For the induction step, we assume that the statement is true when $N=s$, that is, $P(G_s|T,v)=k_s\cdot e^{-(s(d-2)+2)T}\cdot (e^{(d-2)T}-1)^{s-1}$, where $k_s$ is a number only related to the network structure and $v$ is a leaf of $G_s$. To construct $G_{s+1}$, we replace an uninfected neighbor of $v$ with an infected node, say $u$. Hence, $u$ is a leaf of $G_{s+1}$. By using the induction hypothesis and (\ref{eq:adj_prob_int}), we can compute $P(G_{s+1}|T,u)$ as follows

\begin{align*}
    &P(G_{s+1}|T,u)\\
    &=e^{-(d-1)T}\int_0^T \left[ k_s\cdot e^{-(s(d-2)+1)(T-x)} \right. \\ 
    & \indent \left. \cdot (e^{(d-2)(T-x)}-1)^{s-1} e^{-x}\right]   \mathrm{d}x\\
    &= k_s e^{-((s+1)(d-2)+2)T}\int_0^T (e^{(d-2)(T-x)}-1)^{s-1} e^{s(d-2)x}\mathrm{d}x \\
    &= k_s e^{-((s+1)(d-2)+2)T}\int_0^T \left[ \sum\limits_{i=0}^{s-1} \binom{s-1}{i} \right. \\
    & \indent  \left. \cdot e^{(d-2)[T(s-1-i)+(i+1)x]}\cdot (-1)^i  \right] \mathrm{d}x.
\end{align*} 

The first term $e^{-(d-1)T}$, after the first equation, is from the $d-1$ uninfected neighbors of $u$, since $u$ is a leaf of $G_{s+1}$. The third equation is followed by the binomial expansion. The following is the resultant $P(G_{s+1}|T,u)$ after the integration:

\begin{align*}
    &\frac{k_s}{d-2} e^{-((s+1)(d-2)+2)T} \biggl\{ \sum\limits_{i=0}^{s-1}
    \left[e^{(d-2)Ts} \binom{s-1}{i} \frac{(-1)^i}{i+1}\right]\\
    & \indent  -  \sum\limits_{i=0}^{s-1}
    \left[\binom{s-1}{i} \frac{(-1)^i}{i+1} e^{(d-2)T(s-1-i)}\right] \biggr\}\\
    &=\frac{k_s}{(d-2)s} e^{-((s+1)(d-2)+2)T}(e^{(d-2)T}-1)^s.
\end{align*}
By setting $k_{s+1}=\frac{k_s}{(d-2)s}$, we complete the proof of the first statement. Note that from the above result, we can deduce that $\frac{P(G_{s+1}|v,T)}{P(G_{s+1}|u,T)}=\frac{s}{1}$, which is independent of $T$. Based on the above equation, we can compute the exact likelihood of being the source of any vertex in a degree regular tree in $O(N^2)$.

\subsection{Proof of Lemma~\ref{lem:T_inde}}
\begin{figure}[H]
    \centering
    \includegraphics[scale =0.25]{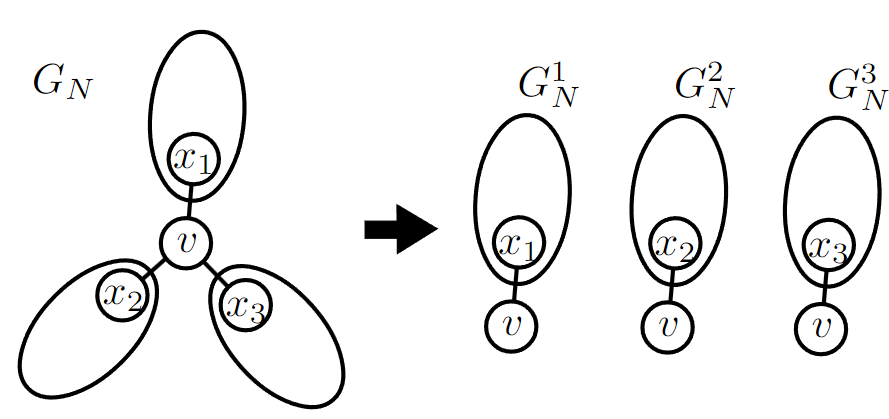}
    \caption{Illustration of how the computation of $P(G_N| v,T)$ of a non-leaf vertex $v$ with three neighboring nodes can be transformed to the multiplication of three leaf cases. Node $v$ is not a leaf in $G_N$, but a leaf in $G^1_N$, $G^2_N$ and $G^3_N$. Hence to compute $P(G_N| v,T)$, we can first decompose $P(G_N| v,T)$ as $P(G^1_N| v,T)\cdot P(G^2_N| v,T) \cdot P(G^3_N| v,T)$ then directly apply the result in leaf node to $P(G^i_N| v,T)$, for $i=1,2,3$.}
    \label{fig:lem_1_leaf}
\end{figure}

We extend the result to other non-leaf nodes. Given a non-leaf vertex $v\in V(G_N)$, without loss of generality, let $x_i\in V(G_N)$ denote infected neighbors of $v$, for $i=1,2,\ldots d$. We consider the likelihood that $v$ being the source in $d$ subtrees of $G_N$, say $P(G^i_N|v,T)$ separately as shown in Fig. \ref{fig:lem_1_leaf}, since they are mutually independent. We have

\begin{align*}
&P(G^i_N|v,T)=\frac{k_ie^{-\left[(|g^v_{x_i}|+1)(d-2)+3-d\right]T}\cdot(e^{(d-2)T}-1)^{|g^v_{x_i}|}}{(d-2)|g^v_{x_i}|}
\end{align*}
Note that in the above $P(v\cup g^v_{x_i}| v,T)$, we have removed $d-1$ uninfected neighbors from $v$. Hence, $v$ has only one neighbor which is infected. We can construct $P(G_N|v,T)$ by combining the above results which leads to 

\begin{align*}
    &P(G_N|v,T)=\prod\limits_{i=i}^d P(v\cup g^v_{x_i}| v,T)\\
    &=k_v\cdot e^{-T \sum\limits_{i=1}^d\left[(|g^v_{x_i}|+1)(d-2)+3-d\right]} \cdot(e^{(d-2)T}-1)^{\sum\limits_{i=1}^d |g^v_{x_i}|} \\
    &= k_v\cdot e^{-(N(d-2)+2)T}(e^{(d-2)T}-1)^{N-1}, 
\end{align*}
where $k_v=\prod\limits_{i=i}^d \frac{k_i}{(d-2)|g^v_{x_i}|}$ and the third equation equation is followed by $\sum\limits_{i=1}^d |g^v_{x_i}|=N-1$.

\subsection{Proof of Theorem~\ref{thm:like_ratio_reg}}

Let $(u,v)\in E(G_N)$ and $x_i, y_i \in V(G)$ are neighbors of $u$ and $v$ respectively for $i=1,2,\ldots d-1$. Since the ratio of the likelihood of being the source between any two nodes is independent from $T$ by Lemma \ref{lem:T_inde}, we omit the term $e^{-(N(d-2)+2)T}\cdot (e^{(d-2)T}-1)^{N-1}$. Let $k_v$ and $k_u$ be the scalar part of the $P(g^u_v|v,T)$ and $P(g^v_u|v,T)$ respectively. We have
\begin{equation*}
    \frac{P(G_N|v,T)}{P(G_N|u,T)}= \frac{k_v\frac{k_u}{|g^v_{x_1}|+|g^v_{x_2}|+\ldots |g^v_{x_{d-1}}|+1}}{k_u\frac{k_v}{|g^u_{y_1}|+|g^u_{y_2}|+\ldots |g^u_{y_{d-1}}|+1}} =\frac{|g^u_v|}{|g^v_u|},
\end{equation*}
where the second equation is followed by $|g^v_{x_i}|=|g^u_{x_i}|$, $|g^u_{y_i}|=|g^v_{y_i}|$, $1+\sum\limits_{i=1}^{d-1}|g^v_{y_i}|= |g^u_v|$, and $1+\sum\limits_{i=1}^{d-1}|g^u_{x_i}|= |g^v_u|$. 


\subsection{Proof of Lemma~\ref{lemma:properties}}

We first treat the right-hand side of (\ref{eq:leaf}) as a function of $T$, denoted as $g(T)$, with fixed $K_{i,j}$ and $d_i$. For brevity, we denote $K_{i,j}$ as $k$ and $d_i$ as $d$. The derivative of $g(T)$ with respect to $T$ can be computed as follows:
\begin{align*}
 &\frac{d}{dT}\int_0^T \frac{t^{k-1}e^{-t}}{(k-1)!}e^{-(T-t)(d-1)} \mathrm{d}t   \\
 =& \frac{e^{T(1-d)}}{(k-1)!}\left(T^{k-1}e^{T(d-2)} -(d-1)\int_0^T t^{k-1}e^{t(d-2)} \mathrm{d}t \right).
\end{align*}
We can omit the part $\frac{e^{T(1-d)}}{(k-1)!}$ since it is positive for all possible $k$, $d$, and $T$ and thus will not affect the sign of the derivative. Hence, we now consider $g_1(T)=T^{k-1}e^{T(d-2)}$ and
$g_2(T)=(d-1)\int_0^T t^{k-1}e^{t(d-2)}$ separately. Since $v_i$ is a leaf of $G_N$ but not a leaf of $G$, we have $d\geq 2$, and thus both $g_1(T)$ and $g_2(T)$ are strictly increasing functions. Moreover, we have $g_1'(T)\geq g_2'(T)$ for $T<k-1$, and $g_1(T)=g_2(T)$ for $T=0$. Hence we can conclude that $g'(T)\geq 0$ initially and there is a $T_{\text{max}}\in  [k-1,\infty)$ such that $g'(T^*)<0$.  

For the second part, we consider the case as $T$ goes to infinity. We have
$$\begin{array}{rl}
\!\!\!\!\displaystyle\lim_{T \to \infty}\frac{\int_0^T t^{k-1}e^{(d-2)t }\mathrm{d}t}{(k-1)!e^{(d-1)T}}=&\!\!\!\! \displaystyle\lim_{T \to \infty}\frac{1}{(d-1)e^T}=0.
\end{array}
$$

Next, we prove the second property by fixing $K_{i,j}$, $T$, and consider the difference of (\ref{eq:leaf}) between the case when $d_i=D+1$ and $d_i=D$. We have:

\begin{align*}
&\int_0^T \frac{t^{k-1}e^{-t}}{(k-1)!}e^{-(T-t)D} \mathrm{d}t -\int_0^T \frac{t^{k-1}e^{-t}}{(k-1)!}e^{-(T-t)(D-1)} \mathrm{d}t \\
=&\int_0^T\frac{t^{K_{ij}-1}e^{-t}}{(K_{ij}-1)!}e^{-(T-t)D}\left(1-e^{(T-t)}\right)\mathrm{d}t \leq 0,
\end{align*}
since $0\leq t\leq T$ and $(1-e^{(T-t)})\leq 0$.

\subsection{Proof of Lemma~\ref{lemma:independen}}
Suppose $v_i$ and $v_j$ are two leaf nodes of the rumor graph $G_N$ given the source $v_n$, and we assume that $v_n$ is the lowest common ancestor of $v_i$ and $v_j$. Then, we have 

\begin{align*}
P(v_i\cap v_j\,|\,v_n)& = \int_0^T \int_0^T \hat{f}_i(v_n,t_1)\hat{f}_j(v_n,t_2) \mathrm{d}t_1 \mathrm{d}t_2 \\
&= \int_0^T \hat{f}_i(v_n,t_1)\mathrm{d}t_1 \int_0^T \hat{f}_j(v_n,t_2) \mathrm{d}t_2,
\end{align*}
which implies $P(v_i\cap v_j\,|\,v_n)=P(v_i\,|\,v_n)\cdot P(v_j\,|\,v_n)$.
For the second part of Lemma \ref{lemma:independen}, we assume that $v_n$ is the source and $v_m$ is the lowest common ancestor of $v_i$ and $v_j$, i.e, they are branched from $v_m$. We also assume that the integrand for computing the probability of rumor spreading time from $v_n$ to $v_m$, from $v_m$ to $v_i$, and from $v_m$ to $v_j$ are $f_m(t)$, $\hat{f}_j(v_m,t)$ and $\hat{f}_j(v_m,t)$ respectively. By the above result, the rumor spreading from $v_m$ to $v_i$ and from $v_m$ to $v_j$ are independent. Then, we have the following derivations for $P(v_i\cap v_j\,|\,v_n)$:

\begin{align*}
&P(v_i\cap v_j\,|\,v_n)\\
=& \int_0^T \int_0^{T-t} \int_0^{T-t}  f_m(t)\hat{f}_i(v_m,t_1)\hat{f}_j(v_m,t_2) \mathrm{d}t_1\mathrm{d}t_2\mathrm{d}t\\
=& \int_0^T f_m(t)\left( \int_0^{T-t}  \hat{f}_i(v_m,t_1)\mathrm{d}t_1\int_0^{T-t} \hat{f}_j(v_m,t_2) \mathrm{d}t_2\right)\mathrm{d}t\\
=& \int_0^T \hat{F}_i(t)\cdot\hat{F}_j(t) \mathrm{d}F_m(t)\\
=&\int_0^{F_m(T)} \hat{F}_i(F_m^{-1}(z))\cdot\hat{F}_j(F_m^{-1}(z)) \mathrm{d}z\\
\geq& \frac{1}{F_m(T)}\int_0^{F_m(T)} \hat{F}_i(F_m^{-1}(z)) \mathrm{d}z 
\cdot \int_0^{F_m(T)}\hat{F}_j(F_m^{-1}(z))\mathrm{d}z\\
\geq& \int_0^{F_m(T)} \hat{F}_i(F_m^{-1}(z)) \mathrm{d}z 
\cdot \int_0^{F_m(T)}\hat{F}_j(F_m^{-1}(z))\mathrm{d}z,
\end{align*}
where $z=F_m(t)$ is the cumulative distribution function corresponding to $f_m(t)$, moreover, $\hat{F}_i(t)$ and $\hat{F}_i(t)$ are the resultant functions of $t$ from the two inner integrals. The first inequality is followed by the Chebyshev integral inequality \cite{fink_cheby} since  $\hat{f}_i(v_m,T-F^{-1}_m(z))$ and $\hat{f}_j(v_m,T-F^{-1}_m(z))$ have the same monotonicity in the range of integration according to Lemma \ref{lemma:properties}. The second inequality is then followed by the fact that $F_m(T)\leq 1$. By changing the variable back to $t$, the second part of Lemma~\ref{lemma:independen} is proved.

\subsection{Proof of Proposition~\ref{prop:two}}
Initially, we assume that the nodes in the $2$-regular tree $G_N$ are sequentially indexed as $v_1, v_2, \dots, v_N$. There are only two leaf nodes in a $2$-regular tree, with each leaf node connecting to one susceptible child node, the likelihood of being the source for node $v_i \in V(G_N)$ can be simplified as
$$P(G_N\,|\,v_i,T)=\frac{T^{i-1}e^{-T}}{(i-1)!}\frac{T^{N-i}e^{-T}}{(N-i)!}.$$
Hence, to maximize $P(G_N\,|\,v_i,T)$ we have 
\begin{align*}
\arg\displaystyle\max_{v_i\in V(G_N)}P(G_N\,|\,v_i,T) = \arg\displaystyle\min_{v_i\in V(G_N)}(i-1)!(N-i)!,
\end{align*}
that is, $\hat v = v_{N/2}$ if $N$ is even; and $\hat v = v_{N/2-1}$ or $\hat v = v_{N/2+1}$ if $N$ is odd. 
For the second part of the Proposition \ref{prop:two}, without loss of generality, we assume that $v_e=v_1$, i.e., $v_1$ is a leaf of both $G$ and $G_N$, and we consider the likelihood ratio between $v_i$ and $v_{i+1}$ where $1\leq i \leq N-1$. The likelihood ratio between two nodes is then given as follows:
\begin{align*}
\displaystyle\lim_{T \to \infty} \frac{P(\bar{G}_N\,|\,v_i,T)}{P(G_N\,|\,v_m,T)}&=\displaystyle\lim_{T \to \infty}\frac{\int^T_0 \frac{t^{i-2}e^{-t}}{(i-2)!}\mathrm{d}t \cdot \frac{T^{N-i}}{(N-i)!}}{\int^T_0 \frac{t^{i-1}e^{-t}}{(i-1)!}\mathrm{d}t\cdot \frac{T^{N-i-1}}{(N-i-1)!}}  \\
&\approx \displaystyle\lim_{T \to \infty} \frac{T}{N-i} >0.
\end{align*}
The last approximation is given by the fact that $\int^{\infty}_0 t^{i-2}e^{-t}\mathrm{d}t=\Gamma(i-1)$ and $\int^{\infty}_0 t^{i-1}e^{-t}\mathrm{d}t=\Gamma(i)$\cite{gamma}. The above result shows that as $T$ increases to infinity when $G_N$ is fixed, $v_1$ is the estimated source of $G_N$.

\bibliographystyle{IEEEtran}
\bibliography{main}

\end{document}